\documentclass[twocolumn,epjc3]{svjour3}

\usepackage{float}
\usepackage[OT1]{fontenc} %

\usepackage{longtable}
\usepackage{multirow}
\usepackage[font=small,skip=0pt]{caption}
\usepackage[labelformat=simple]{subcaption}

\usepackage{amsmath}
\usepackage[nodayofweek]{datetime}
\usepackage{enumerate}
\usepackage{xspace}
\usepackage{xcolor}
\usepackage{graphicx}
\usepackage[export]{adjustbox}
\usepackage{booktabs}
\usepackage{hyperref}

\usepackage{algorithm}
\usepackage{algpseudocode}
\newcommand{\bea}{\begin{eqnarray}\begin{aligned}}
\newcommand{\eea}{\end{aligned}\end{eqnarray}}
\RequirePackage[backend=biber,giveninits=true,doi=false,sorting=none,style=numeric-comp]{biblatex}
\usepackage[bitstream-charter]{mathdesign}
\DeclareSymbolFont{usualmathcal}{OMS}{cmsy}{m}{n}
\DeclareSymbolFontAlphabet{\mathcal}{usualmathcal}

\newcommand{\pt}{\ensuremath{p_{\mathrm{T}}}\xspace}

\newcommand{\pcjedi}{\mbox{PC-JeDi}\xspace}

\newcommand{\epcjedi}{\mbox{EPiC-JeDi}\xspace}
\newcommand{\epcgan}{\mbox{EPiC-GAN}\xspace}
\newcommand{\epcfm}{\mbox{EPiC-FM}\xspace}
\newcommand{\mpgan}{\mbox{MP-GAN}\xspace}

\newcommand{\nvidia}{\mbox{NVIDIA\textsuperscript{\textregistered}}\xspace}

\definecolor{airforceblue}{rgb}{0.36, 0.54, 0.66}
\definecolor{darkpurple}{rgb}{0.5, 0.2, 0.8}
\definecolor{dartmouthgreen}{rgb}{0.05, 0.5, 0.06}
\definecolor{taborange}{rgb}{1.0, 0.49, 0.054}
\definecolor{amaranth}{rgb}{0.9, 0.17, 0.31}

\journalname{Eur. Phys. J. C}

\begin{document}

\title{EPiC-ly Fast Particle Cloud Generation 
with Flow-Matching \\
and Diffusion}

\author{Erik~Buhmann\thanksref{addr1} \and
Cedric~Ewen\thanksref{addr1} \and
Darius~A.~Faroughy\thanksref{addr2} \and
Tobias~Golling\thanksref{addr3} \and
Gregor~Kasieczka\thanksref{addr1,addr4} \and
Matthew~Leigh\thanksref{addr3} \and
Guillaume~Quétant\thanksref{addr3} \and
John~Andrew~Raine\thanksref{addr3} \and
Debajyoti~Sengupta\thanksref{addr3} \and
and~David~Shih\thanksref{addr2}
}

\institute{Institut für Experimentalphysik, Universität Hamburg, 22176 Hamburg, Germany\label{addr1} \and
NHETC, Department of Physics and Astronomy, Rutgers University, Piscataway, NJ 08854, USA\label{addr2} \and
Département de physique nucléaire et corpusculaire, University of Geneva, Switzerland\label{addr3} \and
Center for Data and Computing in Natural Sciences (CDCS), Universität Hamburg, 22176 Hamburg, Hamburg, Germany\label{addr4}}

\date{Received: date / Accepted: date}

\maketitle

\begin{abstract}
    
Jets at the LHC, typically consisting of a large number of highly correlated particles, 
are a fascinating laboratory for deep generative modeling.
In this paper, we present two novel methods that generate LHC jets as point clouds efficiently and accurately. We introduce \epcjedi, which combines score-matching diffusion models with the Equivariant Point Cloud (EPiC) architecture based on the deep sets framework. This model offers a much faster alternative to previous transformer-based diffusion models without reducing the quality of the generated jets. In addition, we introduce \epcfm, the first permutation equivariant continuous normalizing flow (CNF) for particle cloud generation. This model is trained with {\it flow-matching}, a scalable and easy-to-train objective based on optimal transport that directly regresses the vector fields connecting the Gaussian noise prior to the data distribution. Our experiments demonstrate that \epcjedi and \epcfm both achieve state-of-the-art performance on the top-quark JetNet datasets whilst maintaining fast generation speed. Most notably, we find that the \epcfm model consistently outperforms all the  other generative models considered here across every metric. Finally, we also introduce two new particle cloud performance metrics: the first based on the Kullback-Leibler divergence between feature distributions, the second is the negative log-posterior of a multi-model ParticleNet classifier.
\end{abstract}

\section{Introduction}
Operating at the energy and intensity frontier, the Large Hadron Collider~(LHC)~\cite{Evans:2008zzb} produces an incredible amount of data from proton-proton and heavy ion collisions which are recorded by several experiments, such as ATLAS~\cite{Aad:2008zzm} and CMS~\cite{CMS-TDR-08-001}.
In order to compare experimental data to theoretical predictions, even larger numbers of simulated events are required.
However, as the data collected by the experiments grows ever larger, the computational resources required for detailed simulation increases.
As such, a key area of focus for future development is reducing the computation requirements for simulation.

In recent years, a wide range of machine learning approaches have been developed, for example with fast surrogate models for detector simulation~\cite{de_Oliveira_2016,Paganini:2017dwg,Paganini:2017hrr,Erdmann:2018jxd,Belayneh:2019vyx,Buhmann:2020pmy,Krause:2021ilc,Krause:2021wez,atlfast3,ATLAS:2022jhk,Adelmann:2022ozp,Krause:2022jna,AbhishekAbhishek:2022wby,Liu:2023lnn,Diefenbacher:2023vsw,Diefenbacher:2023prl,Buckley:2023rez,Hashemi:2023ruu,Dubinski:2023fsy,Acosta:2023zik,Amram:2023onf,Erdmann:2023ngr,Pang:2023wfx,xu2023generative_detector_response_wiht_Cond_NomalizingFlows,
schnake2022_pointFlow} or event generation~\cite{Otten:2019hhl,Hashemi:2019fkn,DiSipio:2019imz,Butter:2019cae,ArjonaMartinez:2019ahl,Gao:2020zvv,Alanazi:2020klf,Bellagente:2020piv,Velasco:2020nqr,Butter:2020tvl,Howard:2021pos,Quetant:2021hgi} which take advantage of cutting edge deep generative models.
These models can further be used for amplifying statistics~\cite{Butter2020qhk, Calomplification}, or modifying underlying distributions over events~\cite{Lin:2019htn,Algren:2023qnb}. %

At the LHC, the vast majority of particles produced in collisions are grouped together and reconstructed as jets, a collimated shower of electrically charged and neutral constituent particles. As high-dimensional objects with highly complex correlations and substructure, jets are extremely interesting objects of study, both from a physics point of view, as a window into perturbative and non-perturbative QCD, as well as from a machine learning (ML) point of view, where they can be targets for basically any ML task (classification, regression, generative modeling, anomaly detection, etc.). 

Recently, there has been much activity in studying jets as a laboratory for deep generative modeling techniques. In particular the permutation invariance of jet constituents leads to a useful representation of jets as generalized point clouds, and it has been an interesting challenge to develop generative models that respect this permutation invariance~\cite{MPGAN,EPiCGAN, PCJedi, FPCD, käch2023attention_MDMA, JetGPT,CaloClouds, PCDroid, CaloClouds_2}. 
In two contrasting examples, PC-JeDi ~\cite{PCJedi} uses self-attention transformer layers in a diffusion framework~\cite{sohldickstein2015deep, song2020generative_estimatingGradients, song2020improved_technieques_for_sorebased_geneerative, ho2020denoising, song2021scorebased_generativemodelling}  which means large jets take a considerable time to generate; while EPiC-GAN \cite{EPiCGAN} employs the framework of generative adversarial networks together with specially designed equivariant point cloud (EPiC) layers whose computational costs scale linearly with the point cloud size --- as opposed to the quadratic scaling for self-attention transformer layers.

So far, normalizing flows --- despite proving to be very powerful and promising in related generative modeling tasks such as fast calorimeter simulation~\cite{Krause:2021ilc, Krause:2021wez, schnake2022_pointFlow, Krause:2022jna, Diefenbacher:2023vsw, xu2023generative_detector_response_wiht_Cond_NomalizingFlows, Buckley:2023rez,Pang:2023wfx} and event generation~\cite{Gao:2020zvv,Bellagente:2020piv}
 --- have not yet explicitly played a major role in the realm of jet cloud generation. One reason could be that normalizing flows are difficult to make permutation-equivariant due to their extremely constrained structure, which requires bijective maps with tractable Jacobians. The authors of \cite{JetFlow} apply ordinary normalizing flows to jet cloud generation without imposing permutation equivariance, which requires explicit additional conditioning on the jet mass to achieve non-trivial fidelity.
 Continuous normalizing flows (CNFs) \cite{chen2019neural_CNF} could be a way around these constraints, and seem suited to set data \cite{satorras2021en,yang2019pointflow}, but training them has been  computationally prohibitive in the past.

In this paper, we advance point cloud generation of jets in two major directions.
First, we present a combination of EPiC layers with the DDIM score matching approach used in Ref.~\cite{PCJedi}, which we dub EPiC-Jedi, in order to make it both fast and accurate. 
Second, we introduce the first-ever jet cloud generation framework trained using the flow-matching~(FM) paradigm ~\cite{liu2022flow_flowmatching1, albergo2023building_flowmatching2, lipman2023flow_flowmatching3}. (For a previous application of flow-matching to the realm of {\it event generation} -- generating the four vectors of jets instead of their constituents -- see \cite{JetGPT}.)
Combined with an EPiC layer we achieve EPiC-FM, a fast and accurate permutation-equivariant jet cloud generator based on continuous normalizing flows. As both models share the same architecture and can be sampled with the same ordinary differential equation (ODE) solvers, we are able to directly compare these two generative paradigms.

Additionally, we investigate the impact of conditional generation vs. unconditional generation. 
\epcgan is an unconditional generative model, while \pcjedi is a conditional generative model using jet transverse momentum  ($p_\mathrm{T}$) and jet mass as conditioning variables. 
For \epcjedi and \epcfm we train both a conditional and an unconditional version.
We benchmark the performance of the \epcjedi and \epcfm models -- for both unconditional and conditional generation -- and find overall comparable model performance with the highest generative fidelity achieved by  the conditional \epcfm model.

Along the way, we will present a unified overview of diffusion models and continuous normalizing flows, following the ML literature. We will review how both diffusion models and continuous normalizing flows can be understood as two expressions of the same family of generative models, which, for lack of a better name, we will dub {\it continuous-time generative models}. We will see how flow-matching is in some sense the more general paradigm, and in the special case of Gaussian probability paths becomes equivalent to the score-matching objective more commonly used in diffusion models. Finally, we will show how (again, for Gaussian probability paths), the same model can be sampled with a stochastic differential equation (as is customary for diffusion models) or with an ordinary, deterministic differential equation (as is customary for continuous normalizing flows). 
So with this understanding, all the models considered in this work (PC-JeDi, EPiC-JeDi and EPiC-FM) can be viewed as different realizations of continuous-time generative models, with different choices for training objectives and sampling that are all formally equivalent (i.e.\ amount to different hyperparameter choices).

The approach taken in this work to reduce the computational cost differs from the distillation method applied to the FPCD model in Ref.~\cite{FPCD} as the focus is on optimising the network architecture rather than the training and generation procedure.
This enables comparisons of \epcjedi and \epcfm to \pcjedi which are directly analogous to comparisons of \epcgan to \mpgan \cite{MPGAN}, and factorises the differences in performance arising from the choice of architecture and type of generative model.
Concurrent with this work, Ref.~\cite{PCDroid} utilised a new diffusion parametrisation, observing improved performance over \pcjedi on the same network architecture. This parametrisation shares some similarities to flow-matching for training continuous normalising flows, and future comparisons would be of interest.

The outline of the rest of the paper is as follows.
In Sec.~\ref{sec:generalities} we provide an overview  of both diffusion models and continuous normalizing flows and how they are related to each other.
In Sec.~\ref{sec:generative_models} we introduce the modelling paradigms and the permutation-equivariant architecture used in both \epcjedi and \epcfm.
The jet dataset and evaluation metrics used in our case-study to benchmark the performance of these models are described in Sec.~\ref{sec:jet_generation}.
In Sec.~\ref{sec:results} we present the generative fidelity of our models on various jet observables as well as benchmark the sampling time of the EPiC models compared to \pcjedi. 
We draw our conclusions in the final Sec.~\ref{sec:conclusion}.

\section{Continuous-Time Generative Models}
\label{sec:generalities}

The underlying idea behind both flow-based and diffusion-based generative modeling is to evolve a simple data distribution, such as a standard Gaussian, into a more complex distribution that approximates the training data. Originally, normalizing flows achieved
this through a series of discrete, invertible transformations that are parameterized by a deep neural network. 
Recently there has been considerable interest in a potentially
more expressive and flexible class of generative models characterized by {\it continuous-time dynamics} that smoothly evolve the data distribution across an auxiliary time variable. Continuous normalizing flows (CNF)  and diffusion models are prime examples of such models. 

In this section, we will provide a brief overview of various continuous-time generative modeling\footnote{The terminology of ``continuous-time generative modeling" is not standard in the ML literature as far as we know, but we feel it is an appropriate umbrella term that encompasses all the approaches that are being considered at present, both in HEP and more broadly.} approaches and how they relate to one another. In the process, we will aim to clarify some of the relevant terminology and definitions. We will see that flow-matching with CNFs is in some sense the broader framework, encompassing a more general family of probability paths. Diffusion models can be thought of as a special case of flow-matching with  Gaussian probability paths (see Sec.~\ref{sec:gaussian_paths}). Indeed, for Gaussian probability paths, 
we will see that the various approaches to continuous-time generative modeling are distinguished in two principal ways: their training objectives (flow-matching or score-matching) and their sampling (deterministic or stochastic). Score-matching leads to what are usually referred to as diffusion models in the ML literature, but we will review how it is formally equivalent to flow-matching for Gaussian probability paths. So these two training objectives can just be viewed as different hyperparameter choices of the same overall framework.

\subsection{Flow-matching explained}

In a continuous normalizing flow (CNF), one seeks to learn a neural network approximation $v_\theta(x,t)$ to a time-dependent vector field $u_t(x_t)$ that generates a continuous transformation of the data $x_t$:
\begin{equation}\label{eq:ode}
{dx_t\over dt}=u_t(x_t)\,.
\end{equation}
At $t=0$, $x_t$ follows the data distribution $p_{\rm data}(x)$, and at $t=1$ it follows some simple, pre-specified base distribution $p_{\rm base}(x)$. (As is customary, we will take this to be the unit normal distribution for simplicity.)  In between, $x_t$ follows an interpolating distribution $p_t(x)$ which is referred to as the {\it probability path}. This is a continuous time generalization of the ordinary normalizing flow, which implements the map from data to the base distribution through a discrete set of steps.

Continuous normalizing flows have been around for some time, but the computational difficulties of training them with the usual maximum likelihood objective proved to be a significant obstacle to their widespread adoption as generative models for complex, high-dimensional datasets (such as jet clouds and calorimeter showers).

With the advent of flow-matching (FM) \cite{lipman2023flow_flowmatching3}, training CNFs becomes feasible and effective. The idea of flow matching (inspired by diffusion models) was to  attempt to learn the transformation (\ref{eq:ode}) by matching the neural network to a {\it conditional vector field} $u_t(x|x_0)$
via a mean squared error (MSE) loss, instead of the maximum likelihood loss. 
The conditional vector field $u_t(x|x_0)$ generates a {\it conditional probability path} $p_t(x|x_0)$ that interpolates from a delta function\footnote{In practice this delta function is a narrow Gaussian distribution.} $p_t(x|x_0)=\delta(x-x_0)$  at $t=0$, to the ($x_0$-independent) base distribution $p_t(x|x_0)=p_{\rm base}(x)$ at $t=1$.
Integrating this over the data distribution produces an unconditional probability path with the correct boundary conditions
\begin{equation}\label{eq:uncondpt_int}
p_t(x)=\int dx_0\,p_t(x|x_0)p_0(x_0)\,.
\end{equation}
The vector field that generates this probability path is given by an aggregation of the conditional vector field as
\begin{equation}\label{eq:uncondut_int}
u_t(x)=\int dx_0\, u_t(x|x_0){p_t(x|x_0)p_0(x_0)\over p_t(x)}\,.
\end{equation}
The authors of
Ref.~\cite{lipman2023flow_flowmatching3} showed that this follows from applying the continuity equation to both sides of  (\ref{eq:uncondpt_int}).
They further showed that minimizing the loss
\begin{equation}\label{eq:FMloss}
|| v_\theta(x_t,t)-u_t(x_t|x_0)||^2
\end{equation}
over samples $x_0\sim p_0(x)$ drawn from the data distribution, with uniformly sampled times $t\sim{\mathcal U}[0,1]$, and $x_t\sim p_t(x_t|x_0)$, one actually finds $v_\theta(x,t)=u_t(x)$ at the minimum.

\subsection{Gaussian conditional probability paths}
\label{sec:gaussian_paths}

The flow-matching paradigm is quite general and works for any conditional probability path $p_t(x|x_0)$ that goes from a delta function to the base distribution.
One still has to choose a specific path in order to proceed further, however. A natural conditional probability path, which is often assumed in the literature (and will be assumed in all the specific approaches considered in this work) is the {\it Gaussian conditional probability path},
\begin{equation}\label{eq:ptcond_gauss}
p_t(x|x_0)={\mathcal N}(x|\gamma_t x_0,\sigma_t),
\end{equation}
whose trajectories correspond to:
\begin{equation}\label{eq:gaussian_prob_path}
x_t=\gamma_t x_0+\sigma_t \epsilon\,.
\end{equation}
Here $\gamma_t$ and $\sigma_t$ are any differentiable time-dependent functions of $t$ satisfying the boundary conditions  
$(\gamma_0,\sigma_0)=(1,0)$ and $(\gamma_1,\sigma_1)=(0,1)$. This takes a point $\epsilon\sim {\mathcal N}(0,1)$ at $t=1$ and collapses it to $x_0$ at $t=0$. The conditional vector field that generates this transformation is
\begin{equation}\label{eq:utcond}
u_t(x_t|x_0) = {\dot{\sigma}_t\over \sigma_t} x_t + \left({\dot{\gamma}_t\sigma_t-\dot{\sigma}_t\gamma_t\over\sigma_t}\right)x_0 = \dot{\gamma}_tx_0+\dot{\sigma}_t\epsilon
\end{equation}

\subsection{Relation to score-based generative models}

In score-based generative models, Gaussian probability paths (\ref{eq:ptcond_gauss}) are always assumed, but  instead of learning the vector field $u_t(x)$, one learns the score function
\begin{equation}
s_t(x)=\nabla_x\log p_t(x)    
\end{equation}
via matching with the conditional score function instead, analogous to (\ref{eq:FMloss}):
\begin{equation}\label{eq:diffloss}
|| v_\theta(x_t,t)-s_t(x_t|x_0)||^2,
\end{equation}
where
\begin{equation}\label{eq:condscore}
s_t(x_t|x_0)=\nabla_{x_t}\log p_t(x_t|x_0)=-{(x_t-\gamma_t x_0)\over \sigma_t^2} =  -\frac{\epsilon}{\sigma_t}   .
\end{equation}

This can all be seen as a special case of flow-matching, in the following way.  If we plug (\ref{eq:utcond}) and (\ref{eq:ptcond_gauss}) into (\ref{eq:uncondut_int}), we obtain something nice after a bit of calculation:
\begin{equation}\label{eq:utscoreconnection}
u_t(x_t) = f(t)x_t-{1\over2}g(t)^2s_t(x_t),
\end{equation}
with
\begin{equation}\label{eq:fgformulas}
f(t) = {\dot{\gamma}_t\over\gamma_t},\qquad
g(t)=\sqrt{2\sigma_t^2 \left({\dot{\sigma}_t\over\sigma_t}-{\dot{\gamma}_t\over\gamma_t}\right)}.
\end{equation}
So the vector field of a CNF and the score function are actually the same thing when a Gaussian conditional probability path is assumed.\footnote{For non-Gaussian paths there is no relation as far as we are aware, and the CNF vector field is more general.}  In that sense, the MSE losses used in flow-matching and in score-matching can be viewed as different hyperparameter choices and are formally equivalent (in the limit of infinite data, etc). We will refer to these two different training objectives (\ref{eq:FMloss}) and (\ref{eq:diffloss}) as the ``flow-matching objective" and the ``score-matching objective" respectively.

\subsection{Sampling: deterministic (CNF) or stochastic (diffusion)}

With the vector field $u_t(x_t)$ in hand, one can sample deterministically from the ODE (\ref{eq:ode}) using any  numerical ODE solver. One starts with samples from the base distribution as a boundary condition at $t=1$ and then integrates the neural ODE for $x_t$ with respect to $t$, to get new samples from the data distribution.

For the special case of the Gaussian probability path, one has a further option of stochastic sampling, which (in our terminology at least) corresponds to the diffusion model paradigm:
\begin{equation}\label{eq:reverseSDE}
dx_t = \Big[f(t)x_t-g(t)^2s_t(x)\Big]dt+g(t)dw_t
\end{equation}
Here $f(t)$ and $g(t)$ are called the drift and diffusion coefficients and are related to the Gaussian probability path via (\ref{eq:fgformulas}). $w_t$ is the Wiener process and is the source of stochasticity in the sampling. In this case, one solves for $x_t$ (with standard Gaussian boundary conditions at $t=1$) using a numerical SDE solver.

Finally, in the ML literature one also encounters ``diffusion models with deterministic sampling". It was shown in \cite{song2021scorebased_generativemodelling} that for every SDE there is a corresponding ``probability flow ODE" that theoretically gives the same result:
\begin{equation}\label{eq:probflow}
    dx_t = \left[ f(t)x_t - \frac{1}{2}g(t)^2s_t(x_t) \,\right] dt
\end{equation}
This is nothing more than the relation between $u_t$ and the score derived through flow-matching in Ref.~\cite{lipman2023flow_flowmatching3} and reproduced above.
In other words, ``deterministically sampled diffusion models" are really just (special cases of) continuous normalizing flows, and we will refer to them as such. They can be trained using either the flow-matching or the score-matching objective, as explained above.

\section{EPiC-JeDi and EPiC-FM}
\label{sec:generative_models}

In the following we are outlining the specific choices for the score-matching-based training of \epcjedi (Sec.~\ref{sec:epic_jedi})  and the flow-matching-based training of \epcfm  (Sec.~\ref{sec:epic_fm}). 
Additionally, we explain the specific neural network architecture used in both models, the EPiC Network (Sec.~\ref{sec:epic_network}), and outline the normalizing flows used to generate the conditioning vectors (Sec.~\ref{sec:conditional_models}). 

\subsection{EPiC-JeDi}
\label{sec:epic_jedi}

EPiC-JeDi is a permutation-equivariant score-based model for particle clouds. The probability paths are inspired by the diffusion SDE describing the process of adding Gaussian noise to the initial data sample \cite{song2021scorebased_generativemodelling}. Following \cite{PCJedi}, we employ the variance preserving framework \cite{Song2020} where the drift and diffusion are given by $f(t)=-\beta(t)/2$ and $g(t)=\sqrt{\beta(t)}$, respectively. 
The function $\beta(t)$ is chosen such that $\gamma^\mathrm{JeDi}_t$, and $\sigma^\mathrm{JeDi}_t$ follow a cosine schedule, as described in Ref.~\cite{ImprovedDDPM}. 
This leads to the variance preserving path $x_t$ in \eqref{eq:gaussian_prob_path} with:
\begin{align}
    \gamma^\mathrm{JeDi}_t &=\exp \left(- \frac{1}{2} \int_0^t \beta(s) \mathrm{d} s\right),\\
    \sigma^\mathrm{JeDi}_t &=\sqrt{1-(\gamma^\mathrm{JeDi}_t)^2}\,.
\end{align}
 The precise loss function used in EPiC-JeDi is given by the rescaled MSE loss
\begin{align}\label{eq:Diffloss}
        \ell_{\rm JeDi} & \bigl(v_{\!\theta}, s_t(x| x_0)\bigr)\notag \\ & = \left(1+\alpha\frac{\beta(t)}{\sigma(t)^2}\right)\, \Bigl|\Bigl|  v_{\!\theta}(x_t, y, t) - \epsilon\Bigr|\Bigr|^2\,,
\end{align}
with weighting hyperparameter $\alpha$. Consequently, the neural network $ v_{\!\theta}$ is trained to match the noise that was introduced during the diffusion process. We have also included additional contextual information, denoted by $y$, related to the jet-level observables used for the conditional generative models. The rescaling function in front of the MSE loss in \eqref{eq:Diffloss} guarantees that the loss will not diverge when the variance $\sigma_t$ vanishes. In our analysis we fix the weighting parameter to $\alpha=10^{-4}$. 

For sampling new points from the trained \epcjedi model we considered two cases: (i) following Ref.~\cite{PCJedi} we used Euler-Maruyama~(EM)~\cite{EulerMaruyama} SDE solver, and (ii) the Euler and midpoint ODE solvers, a 2\textsuperscript{nd} order Runge-Kutta variant. Notice that for case (i), \epcjedi is a standard score-based diffusion model, while for case (ii) \epcjedi is formally a CNF trained using score-matching.

\subsection{EPiC-FM}
\label{sec:epic_fm}

EPiC-FM is a permutation equivariant CNF trained with the flow-matching objective. Since the flow-matching framework directly regresses any vector-field generating arbitrary paths $x_t$, one could potentially leverage this feature to design or optimize paths that are more linear, or at least less curved, in order to construct simple trajectories between the data and the noise prior. The variance preserving paths in EPiC-JeDi, on the other hand, change non-linearly over time, leading to curved trajectories between data and noise samples. This can translate into less efficient training and sampling; e.g. highly curved trajectories typically require more time-steps when sampling.

For \epcfm we use one of the simplest possible Gaussian probability paths leading to a linear interpolation between the data and the noise sample \cite{lipman2023flow_flowmatching3} \footnote{Note that we are keeping the time flow convention used in diffusion-based models where the data sits at $t=0$ and the Gaussian noise prior at $t=1$. The opposite convention is used in most flow-matching papers.}:
\begin{align}
    \gamma^\mathrm{FM}_t &=1-t,\\
    \sigma^\mathrm{FM}_t &= \sigma_{\text{min}}+(1-\sigma_{\text{min}})t,
\end{align}
with a small minimum noise rate $\sigma_{\text{min}}$ set to $10^{-4}$ and with the conditional vector field now expressed as 
\begin{equation}
    u_t(x|x_0) = (1-\sigma_{\rm min})\epsilon - x_0 \,.
\end{equation}
This particular choice is known as {\it conditional optimal transport} since it leads to straight trajectories with constant-speed that correspond to the optimal transport plan between the conditional probability paths. In this case, the flow-matching loss function is simply given by the MSE loss
\begin{align}
    \ell_{\rm FM} & \bigl( v_{\!\theta}, u_t(x|x_0)\bigr)\notag \\ & = \Bigl|\Bigl| v_{\!\theta}(x_t, y, t) - \bigl((1-\sigma_{\rm min})\epsilon - x_0 \bigr) \Bigr|\Bigr|^2\,.
\end{align}
Here again, we have included in the neural network jet-conditioning information $y$ during training.

We considered the Euler and midpoint ODE solvers for sampling with \epcfm. For a fixed 200 number of function evaluations (NFE)%
\footnote{This leads to different number of time steps between first and second order samplers, instead comparing the performance considering equal computational time.
For the midpoint solver 200 NFE is equal to 100 time steps.}%
, we found the midpoint solver to perform best for both \epcjedi and \epcfm. Hence, for all results generated with \epcjedi and \epcfm in Sec.~\ref{sec:jet_generation} we use the midpoint solver and provide a comparison between the solvers in \ref{app:samplers}.

\subsection {Permutation equivariant architecture}
\label{sec:epic_network}

Both models, \epcjedi and \epcfm, make use of the \textit{equivariant point cloud (EPiC)} layers from Ref.~\cite{EPiCGAN} to parametrize the network $v_\theta(x_t,y,t)$ described in the previous (sub)sections. EPiC layers offer a linear scaling of the computational cost with the number of particles per jet and therefore a faster generation and better scaling to large particle multiplicities than the transformer architecture used in \pcjedi~\cite{PCJedi}.

The architecture is identical between both models and consists of multiple EPiC layers, which are modified to enable conditional generation.
To achieve this,  conditional information is concatenated alongside the input of every multi-layer perceptron (MLP) in the architecture.

\begin{figure*}[h]
    \centering
    \includegraphics[scale=0.8,valign=t]{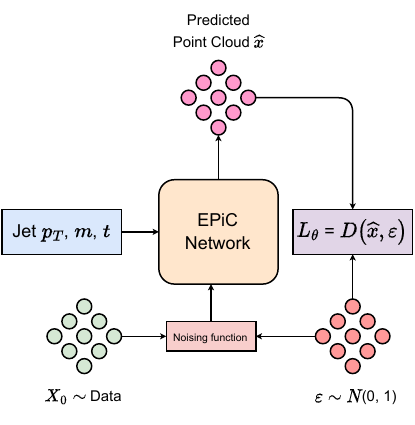}\hskip2ex
    \includegraphics[scale=0.8,valign=t]{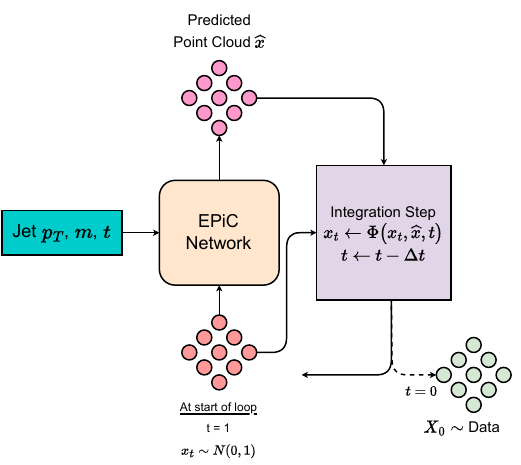}
    \caption{Schematic overview of the \epcjedi and \epcfm training (left) and generation (right) pipeline.
    }
    \label{fig:epicx_scheme}
\end{figure*}

\begin{figure*}[h]
    \centering
        \includegraphics[scale=0.8]{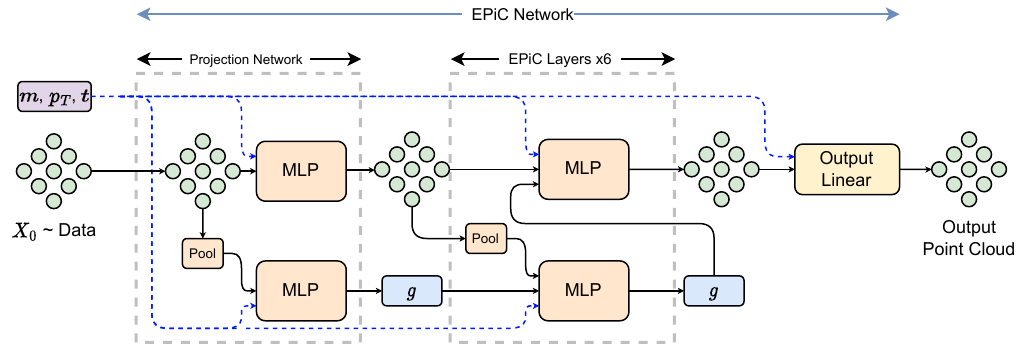}
    \vspace{0.25cm}
    \caption{Model schema of the \textit{EPiC Network} generator architecture used in both \epcjedi and \epcfm. Each multi-layer perceptron (MLP) is a two-layer neural network with LeakyReLU activation. The pooling operation is a concatenation of both average and summation pooling.} 
    \label{fig:model-schema}
\end{figure*}

A schematic overview of the training and generation pipeline for both models is shown in Fig.~\ref{fig:epicx_scheme}. 
The model architecture, dubbed \textit{EPiC Network}, is shown in Fig.~\ref{fig:model-schema}.

The features of the particle cloud used for training and in generation are relative hadronic coordinates --- relative transverse momentum $p_\mathrm{T}^\mathrm{rel}$, relative pseudorapidity $\Delta \eta$, and relative azimuthal angle $\Delta \phi$ --- of the constituents.
The number of constituents is controlled by setting the cardinality of the initial point cloud, i.e. sampling as many noise points as required constituents.
During training, this is always set to the number of constituents of the input jet.
The conditional models are conditioned on the invariant mass and the transverse momentum of the input (target) jet, as well as the current time step.
Their unconditional variants are only conditioned on the current time step.\\

\subsection{Conditional models}
\label{sec:conditional_models}

Due to the correlation between constituent multiplicity and the jet mass, a conditional normalizing flow ~\cite{NormFlows1,NormFlows2,cINNs} is used to generate the joint distribution of jet mass, transverse momentum, and constituent multiplicity during the generation process.
The normalising flow uses four rational quadratic spline (RQS) coupling blocks ~\cite{NeuralSplines} interspersed with invertible linear layers, implemented with the nflows package ~\cite{NFlows}.
Each RQS has 10 bins with linear tail bounds outside $\pm4$.
The conditional normalizing flow transforms samples drawn from a standard multivariate normal distribution into correlated values of jet mass, transverse momentum, and constituent multiplicity. %
The flow is trained to maximise the likelihood of the training set using the standard change of variables formula. 
Dequantisation by adding continuous noise \cite{Dequantisation} is performed on the constituent multiplicity in order to train the normalizing flow, and the nearest integer value is used at generation.

\section{Jet generation}
\label{sec:jet_generation}

We apply \epcjedi and \epcfm to the task of generating top quark initiated jets (top jets) at the LHC.
Following Refs.~\cite{MPGAN,JetFlow,EPiCGAN,PCJedi,FPCD} we use the JetNet datasets~\cite{JetNet,JetNet150} comprising 30 or 150 constituents provided by the JetNet package~(v0.2.2). %
The JetNet data were first introduced under a different name in older work \cite{Coleman:2017fiq,pierini_maurizio_2020_3601436}, studying the impact of calorimeter effects on the performance of jet substructure observables for highly-boosted jets.

Due to their more complex underlying structure, we focus on top jets for the speed and performance comparisons.

\subsection{Dataset}
For the JetNet top-quark datasets, a sample of $\sim200$k $pp\to t\bar t$ events were generated with {\sc MadGraph5} \cite{MG5} at a centre-of-mass energy $\sqrt{s}=$~13~TeV. The transverse momentum for each top-quark was taken inside a narrow window of $\delta p_\mathrm{T}/p_\mathrm{T} =0.01$ around $p_\mathrm{T}=1$ TeV. The hadronic decay, showering and hadronization of the tops were performed with {\sc Pythia8} \cite{Pythia} using the Monash 13 tune \cite{Monash13}, including the underlying event. Subsequently, the event samples were exposed to pileup contamination and  passed through a parametric detector simulation that captures realistic detector effects, such as electromagnetic and hadronic calorimeter granularization and the energy smearing of particles as they propagate through the detector's medium (for more details see \cite{Coleman:2017fiq}). In the last step, the final states in each event were 
clustered into jets with {\sc fastJet} \cite{FastJet} using the anti-$k_t$ algorithm \cite{AntiKt} with a jet radius of $R = 0.4$.\footnote{
As far as we can tell, this radius was incorrectly reported as $R=0.8$ by the original authors of this dataset \cite{Coleman:2017fiq}. This error was then propagated forward in subsequent references, including \cite{MPGAN}. It is clear from the plots of jet $\eta$ and $\phi$, as well as the top jet invariant mass plot which shows a second peak for the $W$, that the jet radius cannot be $R=0.8$. We have also confirmed this by regenerating the JetNet dataset ourselves.} The JetNet dataset is ultimately composed of the leading jet in each event after passing various selection cuts.

The combination of a highly boosted environment with a fairly small clustering radius is what leads to the challenging bimodal feature that can be seen in the jet invariant mass distribution; a small secondary peak located at the $W$-boson mass coming from unresolved top-quarks followed by a much larger resonance located at the value of the top-quark mass corresponding to highly-boosted top-jets. A consequence of parton showering and the detector effects is a significant broadening of most of the momentum-related features, such as the jet $p_\mathrm{T}$ spectrum which becomes much wider than the narrow $p_\mathrm{T}$ selection window of the initial top-quarks.

We allocate $70\%$ of the total data for training and reserve the remaining $30\%$ for the evaluation metrics tests. Detailed information on the architecture, training hyperparameters, and dataset splits is provided in Tab.~\ref{tab:hyperparameters} in Appendix~\ref{app:hyperparams}. This setup allows for a direct comparison with earlier methods like EPiC-GAN and PC-JeDi. 

\subsection{Data representation}

Up to the leading 30 (150) constituents in each jet ordered by decreasing \pt are selected, with all selected in jets with fewer than 30 (150) constituents.
Each constituent is described by its three momentum vector relative to the jet ($\Delta\eta$, $\Delta\phi$, $\pt^\mathrm{rel}$), where
\begin{align*}
\Delta\eta&=\eta^\mathrm{const}-\eta^\mathrm{jet},\\
\Delta\phi&=\phi^\mathrm{const}-\phi^\mathrm{jet},\\
\pt^\mathrm{rel}&=\pt^\mathrm{const}/\pt^\mathrm{jet}.
\end{align*}
The jet four momentum vectors are calculated from the vector sum of all constituents.
All input variables are normalised by their mean and standard deviation in the training dataset.

Furthermore we examine two scenarios for each model:
\begin{itemize}
\item{\bf Unconditional} the models are trained on the input data solely comprising of the jet constituents features $x=(\Delta\eta, \Delta\phi, p_T^{\rm rel})$.

\item{\bf Conditional} the models are trained on jet constituents $x$ conditioned on jet-level features $y=(p_T^{\rm jet}, m^{\rm jet})$. These features have been derived from the data using a normalizing flow paired with a masked autoregressive architecture.
\end{itemize}
Substructure observables are calculated from the jet constituents for the purposes of evaluating the quality of the generative model.
In this work we focus on the N-subjettiness \cite{Thaler_2011} and energy correlation functions~\cite{Larkoski_2013} which are commonly used by the ATLAS and CMS collaborations, as well as the recently introduced energy flow polynomials~(EFPs)~\cite{Komiske_2018}.
To assess the generation performance we follow the procedure introduced in Ref.~\cite{MPGAN} as well as additional measures studied in Refs.~\cite{EPiCGAN,PCJedi}.  
All substructure variables are calculated using the relative \pt of the constituents, and are not renormalised by the inclusive jet \pt.

\subsection{Evaluation metrics}
\label{sec:metrics}

To assess the generation performance of each model, we follow the procedure described in Ref.~\cite{MPGAN} as well as additional measures studied in Refs.~\cite{EPiCGAN,PCJedi}.
However, instead of using the Wasserstein-1 distance between generated showers and the target distributions we measure the agreement using the Kullback-Leibler divergence (KLD).

As the Wasserstein-1 distance in one dimension is calculated as the area between the two cumulative distribution functions, it is very sensitive to overall shifts in distributions.
Its value represents the minimal overall ``work'' required to move a probability distribution to match the target.
This makes it very well suited as a distance measure between to distributions which do not have overlapping support.
However, in the case of similar distributions the sensitivity to an ordered shift makes them less sensitive to the compatibility of the densities of distributions across the full range equally.
An overestimation early in the distribution followed by perfect agreement until an underestimation at the very end will be measured as more dissimilar than an overestimation immediately followed by an equally large underestimation.
In high energy physics, rather than the amount of work required to make two probability distributions match, we primarily consider how well the densities of two samples match at each point in distribution space.
To measure the relative agreement we can instead measure the compatibility between the generated samples and the target distribution using the reverse Kullback-Leibler~(KL) divergence.

A drawback of the KL divergence, in contrast to the Wasserstein-1 distance, is that it requires binned probability distributions and is therefore sensitive to the choice of binning.
For a consistent definition across all samples, we define bin edges using 100 equiprobable quantiles from the target distribution, bounded by the minimum and maximum values.
Any outliers in the reference distributions are subsequently not considered in the calculation of the KL divergences.
This procedure allows us to establish a consistent baseline without introducing a prior bias from the choice of binning and with equivalent statistical sensitivity in each bin.

We further complement these evaluation metrics by introducing a {\it multi-model classifier metric} for generative models~\cite{GalaxyFlow} that generalizes the binary classifier test presented in \cite{Krause:2021ilc}.\footnote{It would have also been interesting to perform the binary classifier test of Ref.~\cite{Krause:2021ilc}, which can quantify the absolute performance of a generative model. However, such method requires a large amount of reference data to properly train the classifier, which the JetNet dataset simply lacks. The multi-model classifier circumvents this issue since it is trained solely on generated events and only requires the reference data for final evaluation.} 
For a specific jet type, an $N$-class classifier is trained to discriminate between the labeled data samples $c = 1, \ldots, N$ generated by $N$ different generative models. After training, this classifier is evaluated on a reference dataset $\mathcal{D}_{\text{test}}$ to compare the performance of the generative models relative to each other. Ideally, the top-performing models will produce a classifier output distribution $p(c|x)$ that aligns closely with the reference data. To gauge this performance, we employ the \textit{negative log-posterior} metric \cite{GalaxyFlow}, expressed as 
\begin{equation}
\text{NLP}(c) \ =\  -\,\mathbb{E}_{x\sim \mathcal{D}_{\text{test}}} \log p(c|x).    
\end{equation}
The generative model $c$ that yields the lowest negative log-posterior is deemed the best. For our classifier, we adopt the ParticleNet architecture \cite{ParticleNet} and employ a cross-entropy loss function. Details about the training hyperparameters can be found in \ref{app:classifier}.

All results presented are calculated from a held out portion of the JetNet dataset with approximately 27,000 jets. 
From each generative model we sample approximately 250,000 events.
The complete available statistics are shown in the Figures in Sec.~\ref{sec:results}.
The KLD scores are calculated using batches of 50,000 test events and generated events sampled with bootstrapping to evaluate the statistical uncertainty.

\section{Results}
\label{sec:results}

In this section, we compare \epcjedi and \epcfm  with the previous state-of-the-art models for conditional and unconditional generation, PC-JeDi and \epcgan, respectively. Since we have observed overall a higher fidelity with the conditional models, we will primarily discuss these results in this section. Additional results for the unconditional models can be found in \ref{app:uncond_results}. The results presented focus on distributions previously shown in work using the same dataset~\cite{EPiCGAN, PCJedi}.

\subsection{Top jets: 30 constituents}
\label{sec:results30}

\begin{table*}[t]
    \centering
    \caption{Top jet generation with 30 constituents: summary of performance metrics for generated jets using the Fr\'echet ParticleNet Distance (FPND), the negative log-posterior (NLP) of a ParticleNet multi-model classifier (with $N\!=\!6$ classes), and the Kullback-Leibler divergence for various features with respect to JetNet-30 test dataset.}
    \label{tab:metrics30}
    \resizebox{\textwidth}{!}{%
    \begin{tabular}{llrrrrrrrrrr}
\toprule
Generation & Model& FPND &  NLP  &$\mathrm{KL}^{m} (\times 10^{-3})$ & $\mathrm{KL}^{p_T^{\rm const}} (\times 10^{-3})$ & $\mathrm{KL}^{\tau_{21}} (\times 10^{-3})$ & $\mathrm{KL}^{\tau_{32}} (\times 10^{-3})$ & $\mathrm{KL}^{D_2} (\times 10^{-3})$ \\

\midrule
\multirow[c]{3}{*}{Conditional} & PC-JeDi & $0.40$ & $3.08$ &$8.56 \pm 0.75$ & $3.25 \pm 0.09$ & $12.82 \pm 1.16$ & $27.08 \pm 1.40$ & $11.91 \pm 0.92$ \\
                                & EPiC-JeDi & $0.42$ & $3.1$ & $5.26 \pm 0.51$ & $2.99 \pm 0.05$ & $\mathbf{7.81 \pm 0.61}$ & $17.34 \pm 1.08$ & $6.58 \pm 0.73$\\
                                & EPiC-FM & $\mathbf{0.11}$ & $\mathbf{1.35}$ & $\mathbf{3.77 \pm 0.50}$ & $\mathbf{2.03 \pm 0.02}$ & $\mathbf{7.40 \pm 0.64}$ & $\mathbf{8.09 \pm 0.93}$ & $\mathbf{4.31 \pm 0.46}$ \\
\midrule
\multirow[c]{3}{*}{Unconditional} & EPiC-GAN & $0.34$ & $3.43$ & $\mathbf{3.71 \pm 0.42}$ & $3.33 \pm 0.03$ & $\mathbf{8.28 \pm 0.76}$ & $17.68 \pm 0.91$ & $13.18 \pm 1.04$ \\
                                  & EPiC-JeDi & $1.63$ & $3.11$ & $18.42 \pm 1.12$ & $3.73 \pm 0.08$ & $\mathbf{8.00 \pm 0.80}$ & $15.27 \pm 1.35$ & $12.33 \pm 1.06$\\
                                  & EPiC-FM & 0.14 & $1.38$ &$5.80 \pm 0.54$ & $\mathbf{ 2.03 \pm 0.01}$ & $\mathbf{7.69 \pm 0.71}$ & $\mathbf{9.24 \pm 1.00}$ & $\mathbf{4.51 \pm 0.58}$\\
\bottomrule
\end{tabular}

    }
\end{table*}

\begin{figure*}[t]
    \centering
        \includegraphics[scale=0.6]{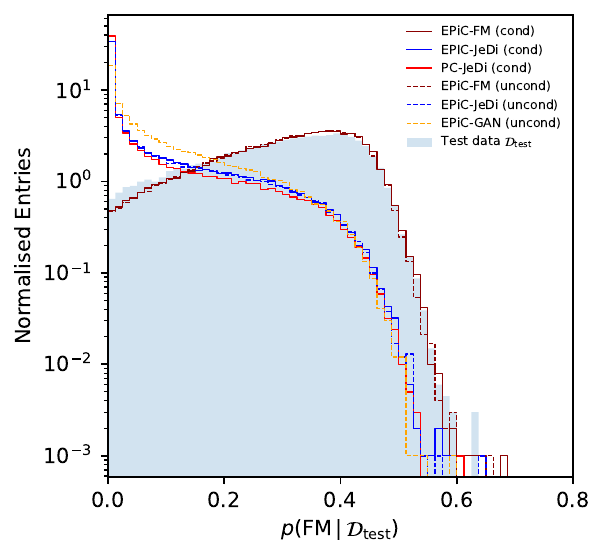}
        \includegraphics[scale=0.6]{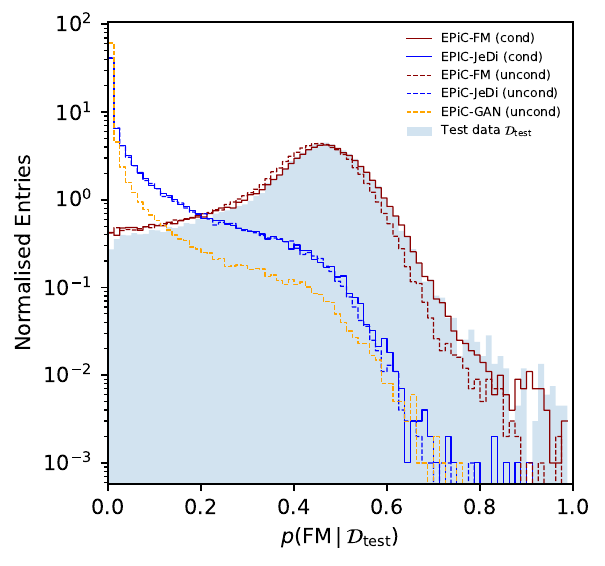}
    \caption{The ParticleNet multi-model classifier output for the conditional EPiC-FM class. We show results for the conditional (solid) and unconditional (dashed) models displayed in Tab~\ref{tab:metrics30} for JetNet-30 (left) and Tab~\ref{tab:metrics150} for JetNet-150 (right). The classifier is evaluated on the test dataset $\mathcal{D_{\rm test}}$ (light blue shade).}
    \label{fig:classifier}
\end{figure*}

\begin{figure*}[ht]
    \begin{center}
    \includegraphics[width=.32\textwidth]{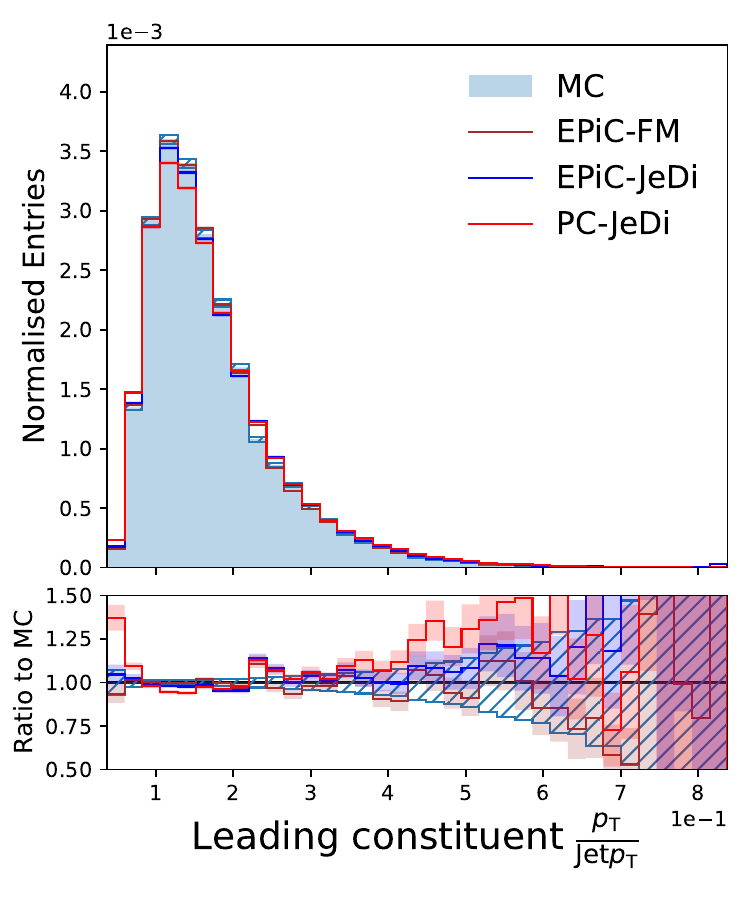}
    \includegraphics[width=.32\textwidth]{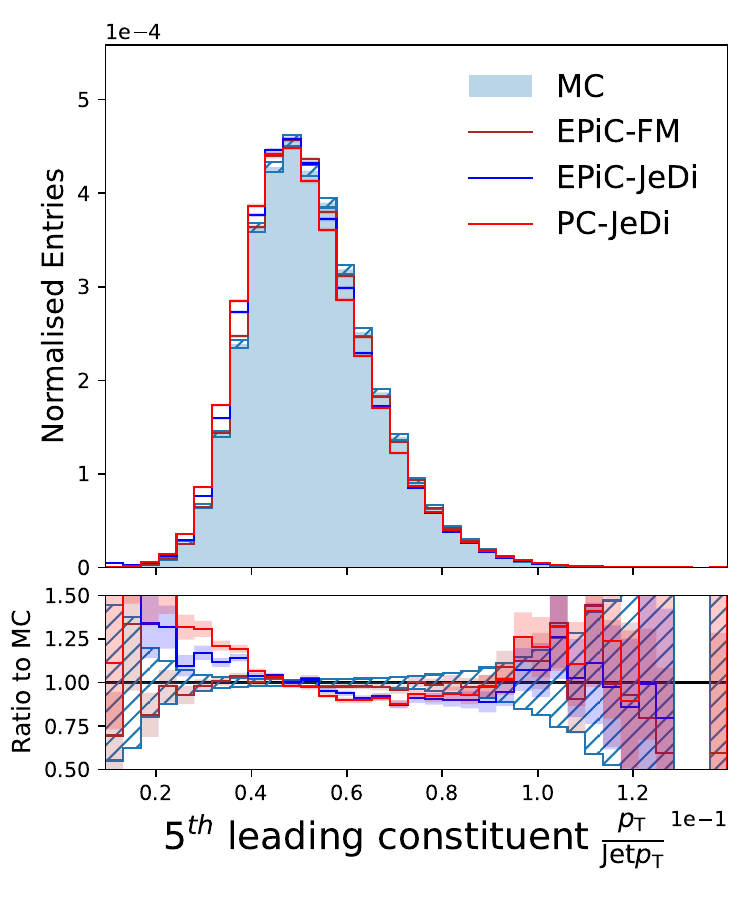}
    \includegraphics[width=.32\textwidth]{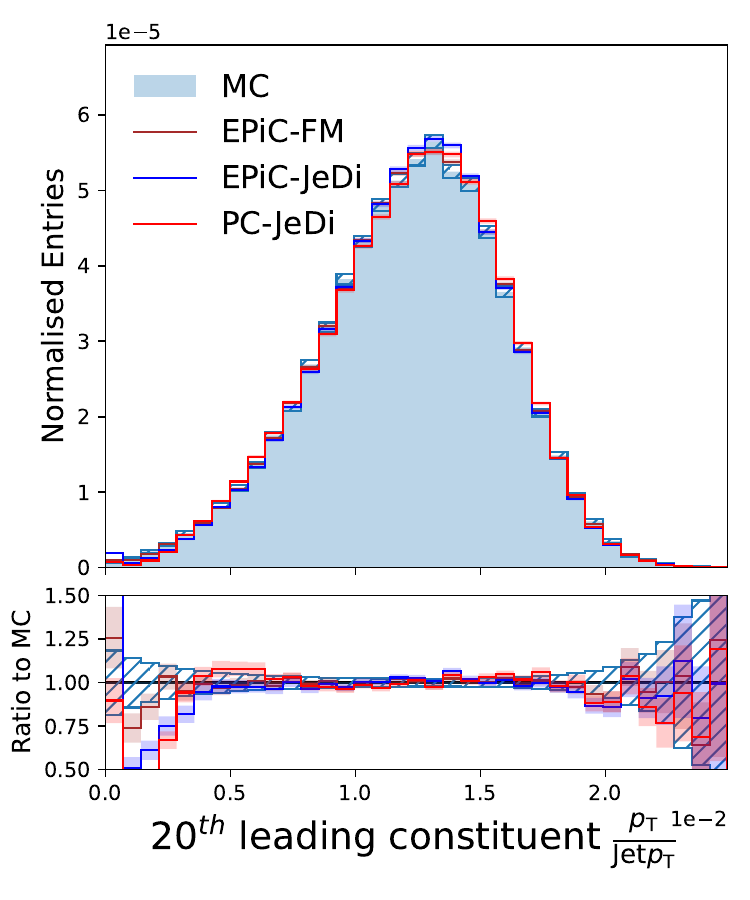}
    \end{center}
    \begin{center}
    \includegraphics[width=.32\textwidth]{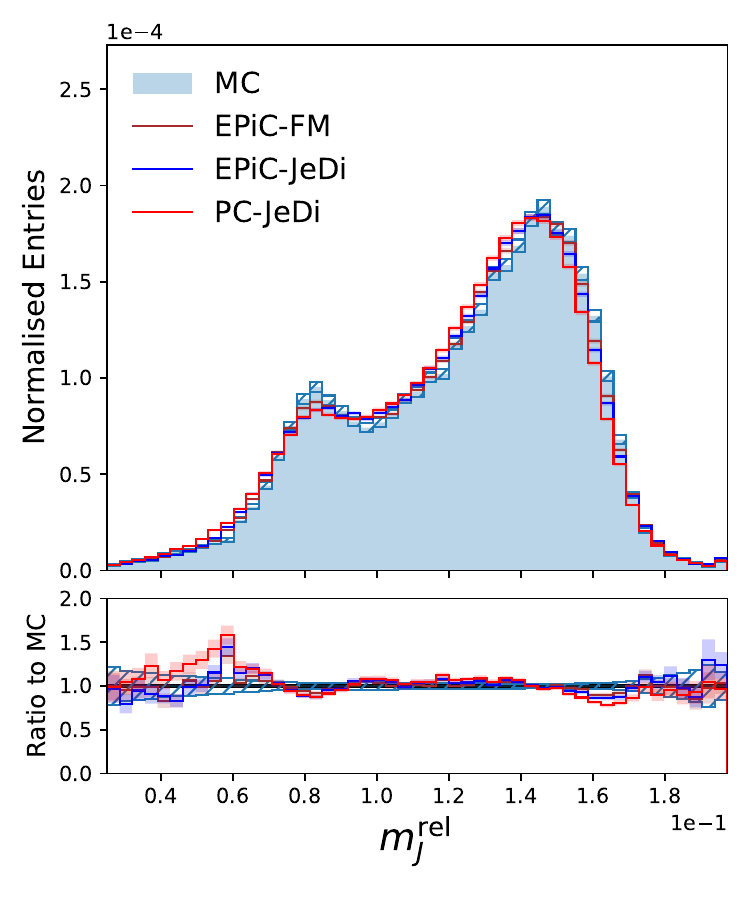}
    \includegraphics[width=.32\textwidth]{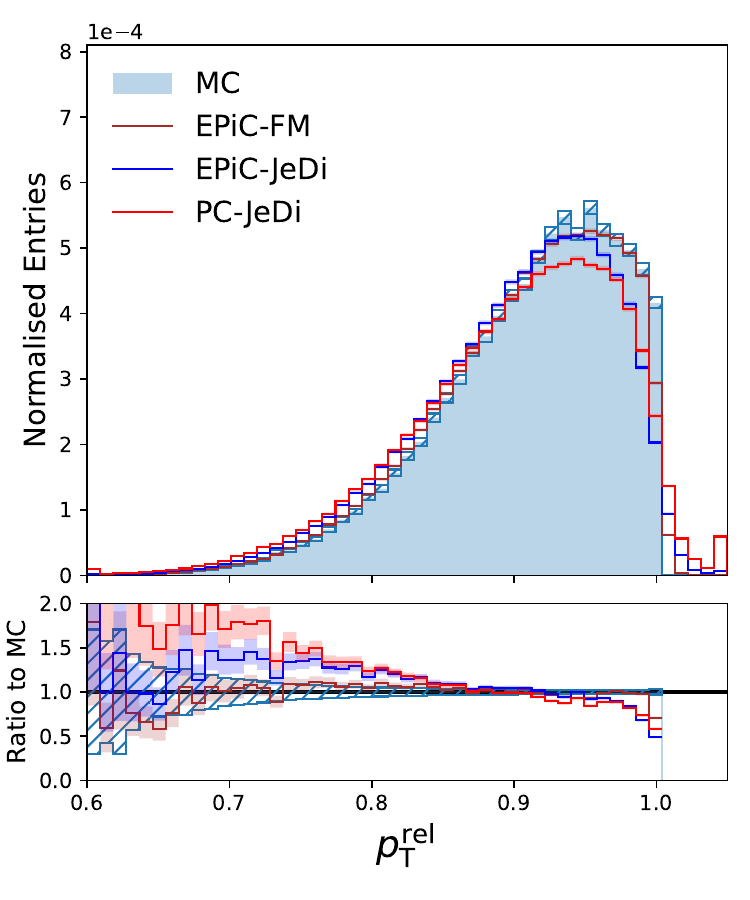}
    \end{center}
    \begin{center}
    \includegraphics[width=.32\textwidth]{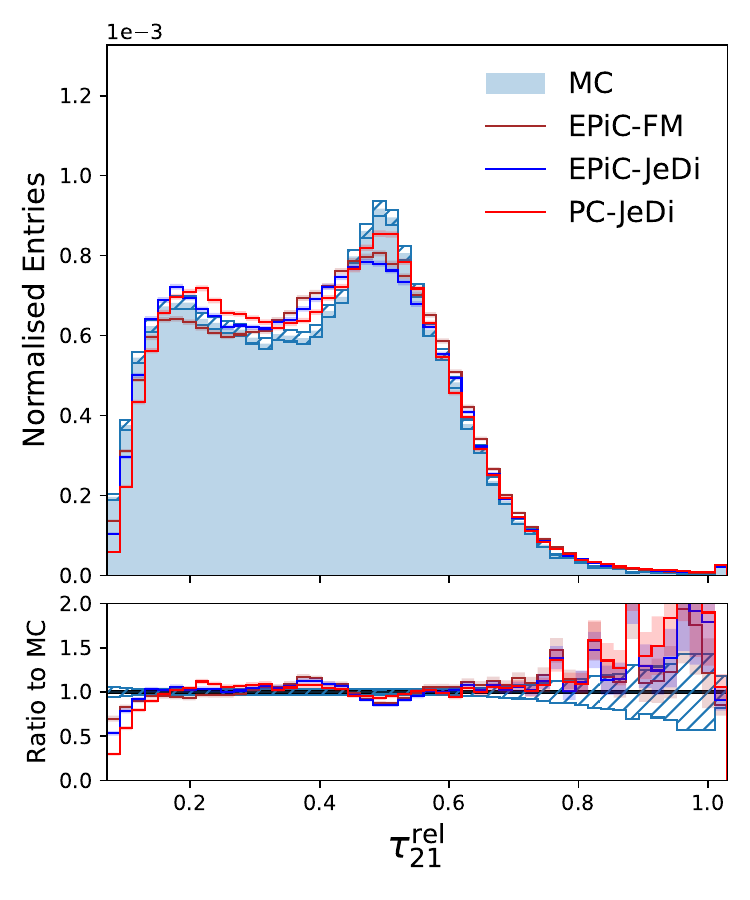}
    \includegraphics[width=.32\textwidth]{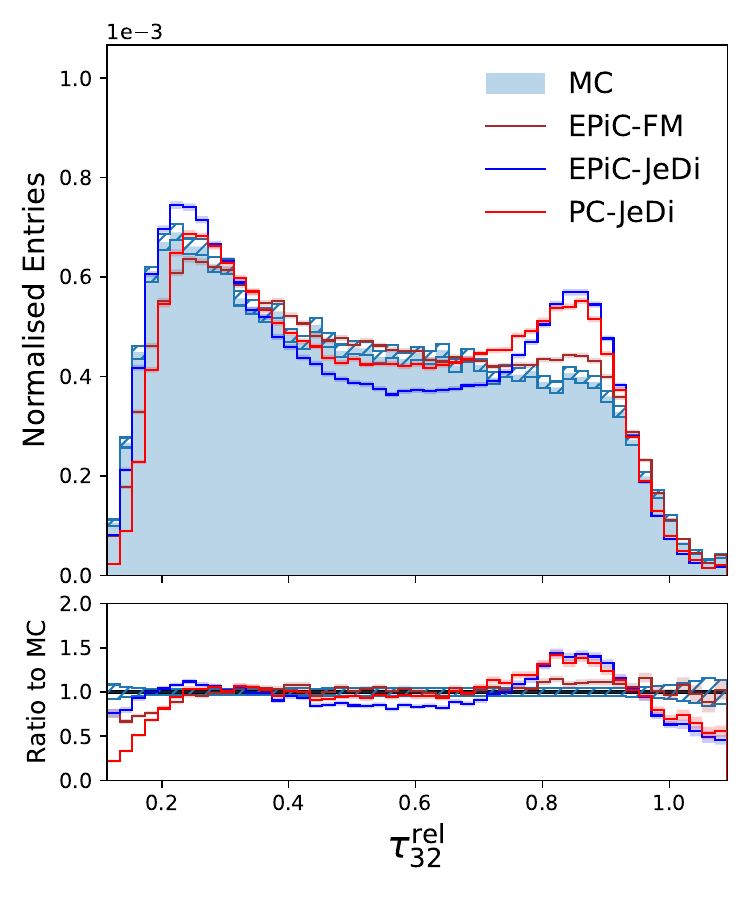}
    \includegraphics[width=.32\textwidth]{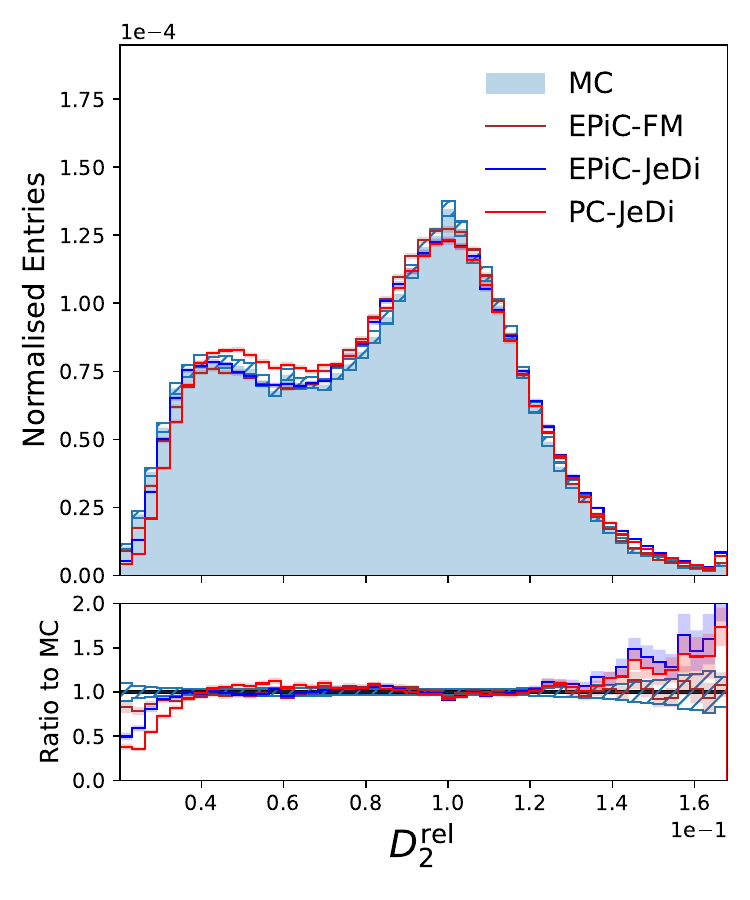} 
    \end{center}
    \caption{Top jet generation with 30 constituents from Monte Carlo (light blue shade) and from conditional EPiC-FM (dark red), conditional EPiC-JeDi (blue) and conditional PC-JeDi (red). First row: $p_T$ distribution of leading (left), fifth leading (middle), and twentieth leading (right) constituents. Second row: distributions of $m^{\mathrm{rel}}$ (left) and $p_T^{\mathrm{rel}}$ (right). Third row: distributions of $\tau_{21}$ (left), $\tau_{32}$ (middle), and $D_{2}$ (right). For EPiC-FM and EPiC-JeDi, a midpoint solver with is used, while for PC-JeDi an Euler Maruyama solver, each with 200 function evaluations. The error bars correspond to statistical uncertainty of the bin count.}
    \label{fig:30-top30}
\end{figure*}

First, we trained both conditional and unconditional EPiC-JeDi and EPiC-FM models on the top-quark JetNet dataset with 30 constituents (JetNet-30). We also train, for comparison, a conditional PC-JeDi model and use the pre-trained version of the unconditional \epcgan model from Ref.~\cite{EPiCGAN}. In Fig.~\ref{fig:30-top30} we show the generated distributions (solid) for the three conditional models, EPiC-FM (dark red), EPiC-JeDi (blue) and PC-JeDi (red) compared to the truth level JetNet-30 samples (blue shade). For EPiC-FM and EPiC-JeDi, we used the midpoint solver, while for PC-JeDi, we used the Euler-Maruyama solver, each with 200 function evaluations. 
All three methods manage to capture very well, within the uncertainties, the relative transverse momentum $p_T^{\rm rel}$ distributions of the leading and subleading constituents. The three models start to deviate from optimal performance when looking at higher level observables that are more sensitive to higher-moments correlations between constituents, like e.g. the jet relative mass $m_j^{\rm rel}$ and relative transverse momentum $p_T^{\rm rel}$ as shown in the middle row of Fig.~\ref{fig:30-top30}. 
This deviation becomes more dramatic when analysing the jet substructure observables such as the $N$-subjetinesss ratios $\tau_{21,32}$ and the energy correlation ratio $D_2$, displayed in the bottom row of Fig.~\ref{fig:30-top30}. Additional results for the unconditional models can be found in see \ref{app:uncond_results}.

To more precisely evaluate these differences, we computed several performance metrics for the conditional and unconditional generative models. The results can be found in Tab.~\ref{tab:metrics30}. The first two columns show the values for the Fréchet ParticleNet distance (FPND) and the negative log-posterior (NLP) derived from the ParticleNet multi-model classifier for $N=6$ models, i.e. trained simultaneously over conditional and unconditional models. Note that these metrics encompass the entire phase space for evaluation. Both metrics agree that the conditional EPiC-FM model is the top-performing generative model, with the unconditional version trailing closely. In Fig.~\ref{fig:classifier} we also present the classifier outputs $p(c={\rm FM}|x)$ for the conditional EPiC-FM class evaluated on the test dataset. From the figure one sees that of all the models, EPiC-FM is the one closest to the test data distribution. Both the FPND and NLP metrics attribute a similar performance to PC-JeDi and EPiC-JeDi. 

The subsequent columns display the  Kullback-Leibler metric KL$^{\rm obs}$ for various high-level jet observables shown in Fig.~\ref{fig:30-top30}  as well as the KL score for the relative constituent $p_\mathrm{T}/\mathrm{Jet}\,p_\mathrm{T}$ distribution. These metrics complement the FPND and multi-model classifier metrics, as they evaluate model performance using one-dimensional phase-space projections onto physically relevant quantities. The results indicate that the conditional EPiC-FM model outperforms other methods and aligns with the findings from the FPND and NLP metrics. The sole exception is the KL$^{m}$ metric for the relative jet mass distribution. For the unconditional models, \epcgan produces a better fit to the truth data, while performing poorly in the rest. This can be visually corroborated in the $m_j^{\rm rel}$ distributions in Fig.~\ref{fig:30-top30uncond} (middle row) where \epcgan (yellow) outperforms the rest. In addition, we also find that \epcgan is outperforming the unconditional \epcjedi for the FPND metric. This apparent superiority in performance can be explained by the model selection procedure that was used during training: for \epcgan, the trained model was specifically selected based on its capability to accurately produce the jet mass, while for all other non-GAN methods the trained models were selected from the last training epoch without any reference to a performance metric.

\clearpage

\begin{figure*}[ht]
    \begin{center}
    \includegraphics[width=.32\textwidth]{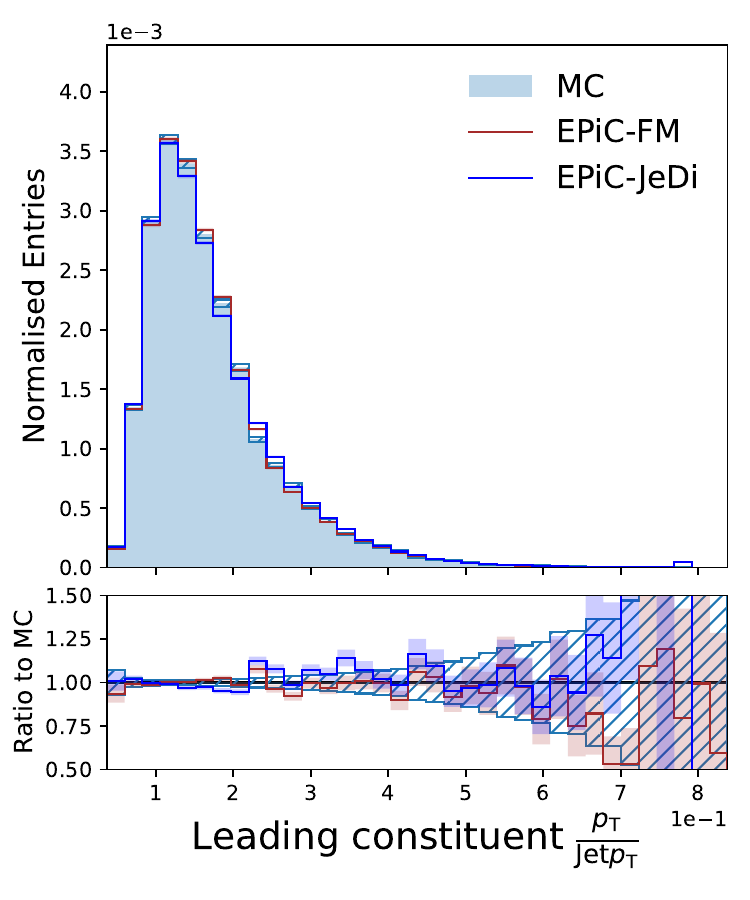}
    \includegraphics[width=.32\textwidth]{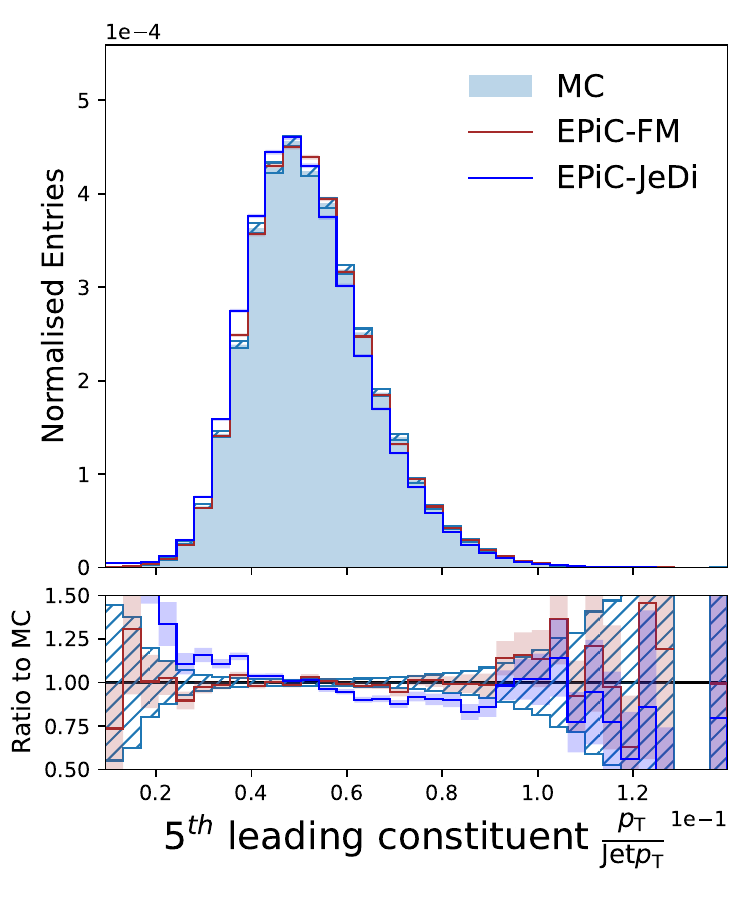}
    \includegraphics[width=.32\textwidth]{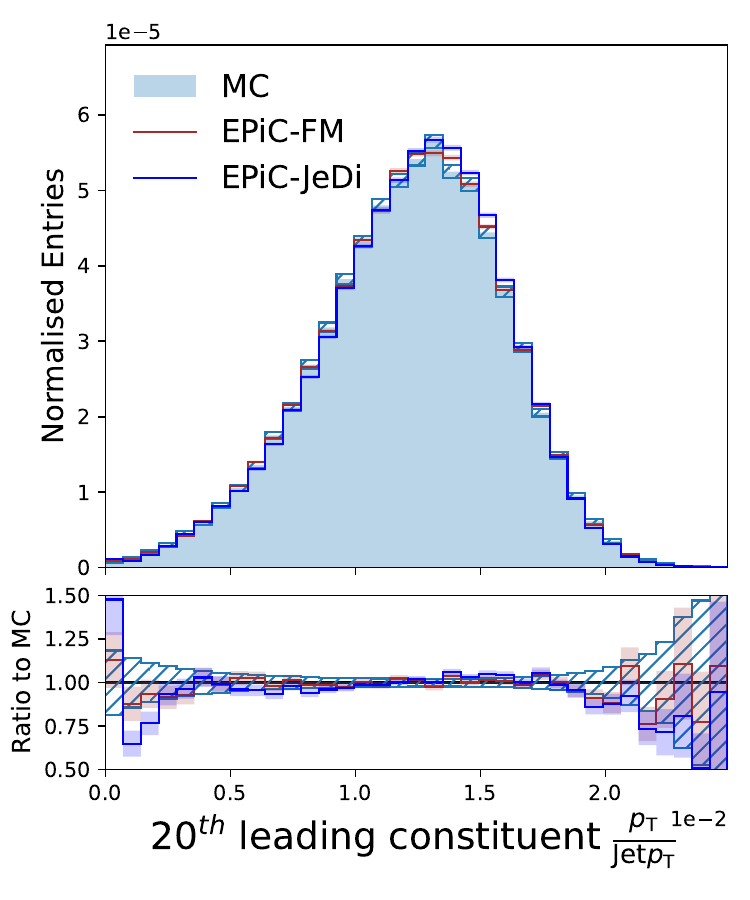}
    \end{center}
    \begin{center}
    \includegraphics[width=.32\textwidth]{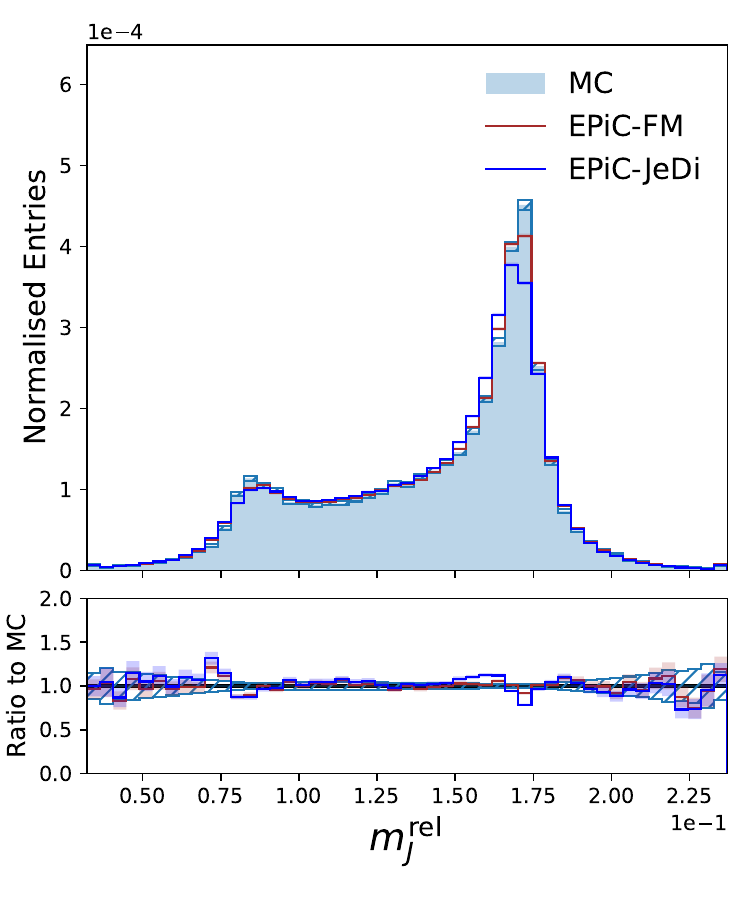}
    \includegraphics[width=.32\textwidth]{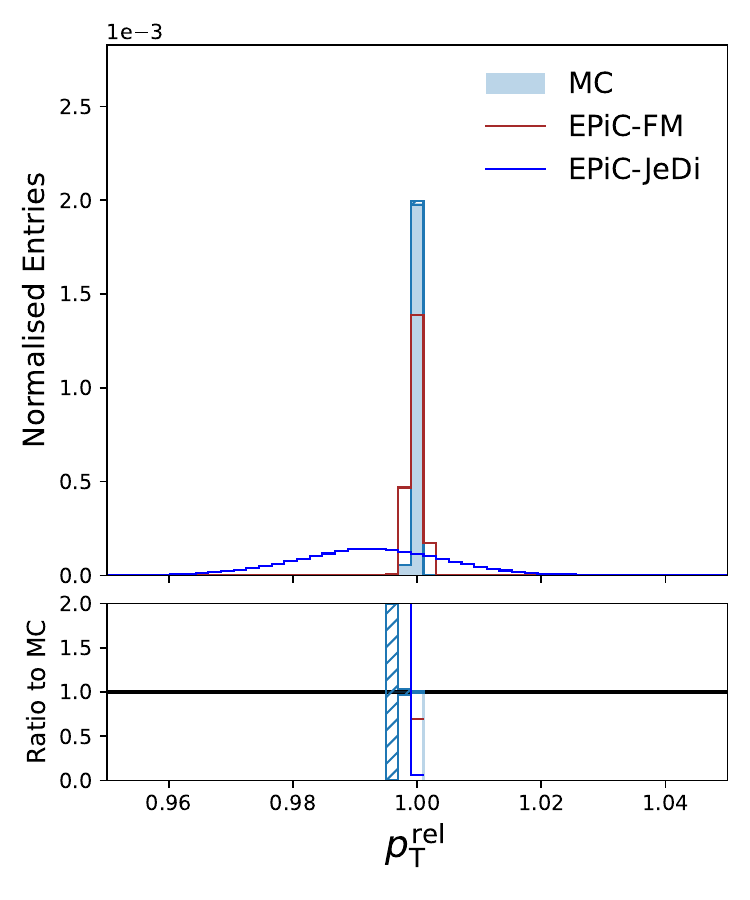}
    \end{center}
    \begin{center}
    \includegraphics[width=.32\textwidth]{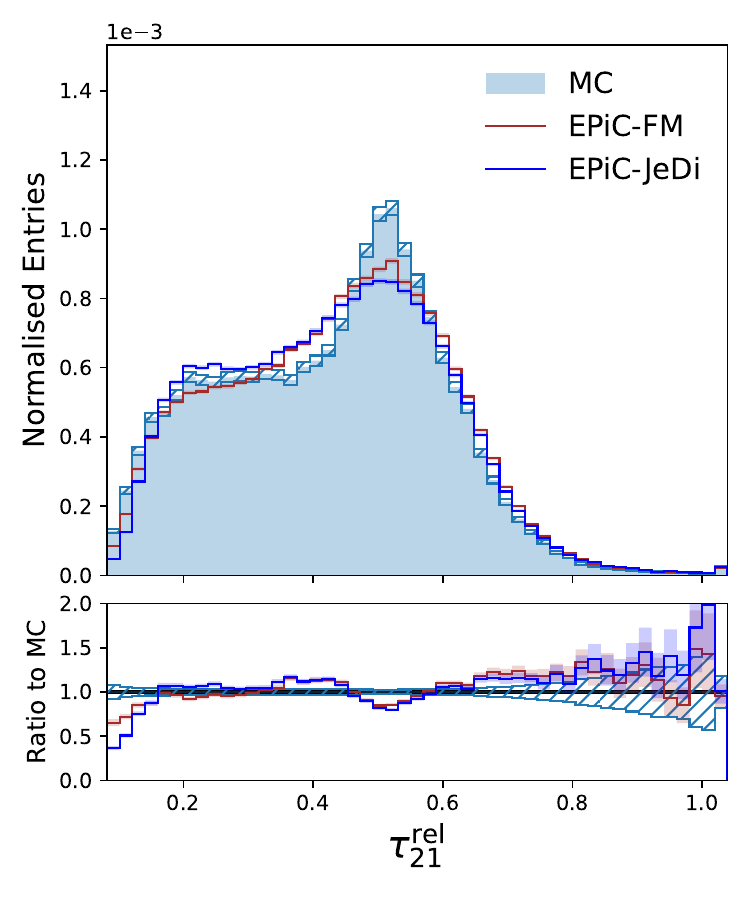}
    \includegraphics[width=.32\textwidth]{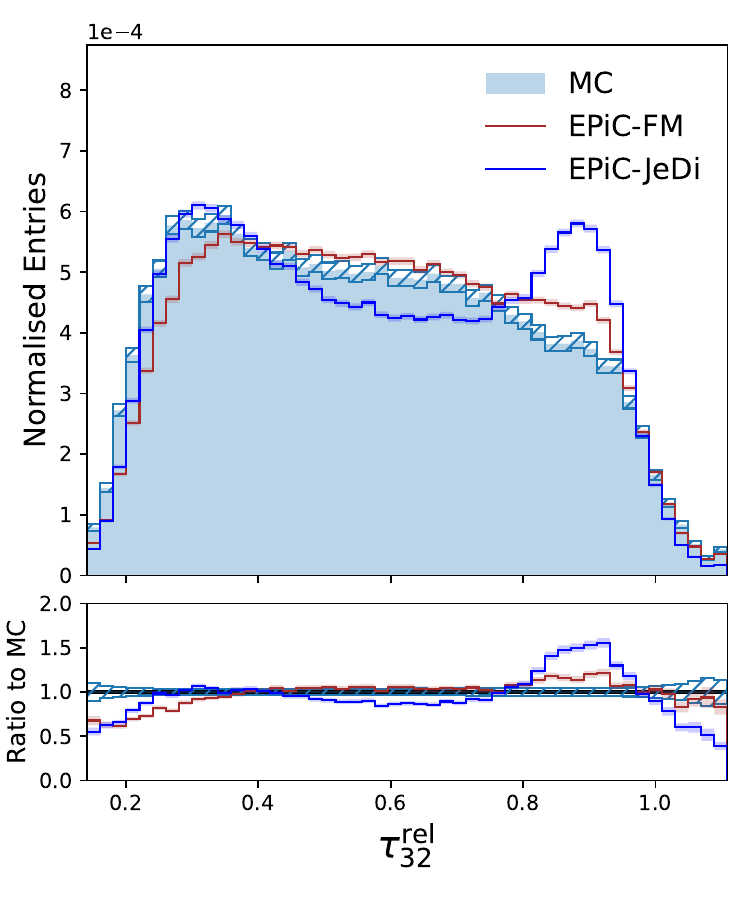}
    \includegraphics[width=.32\textwidth]{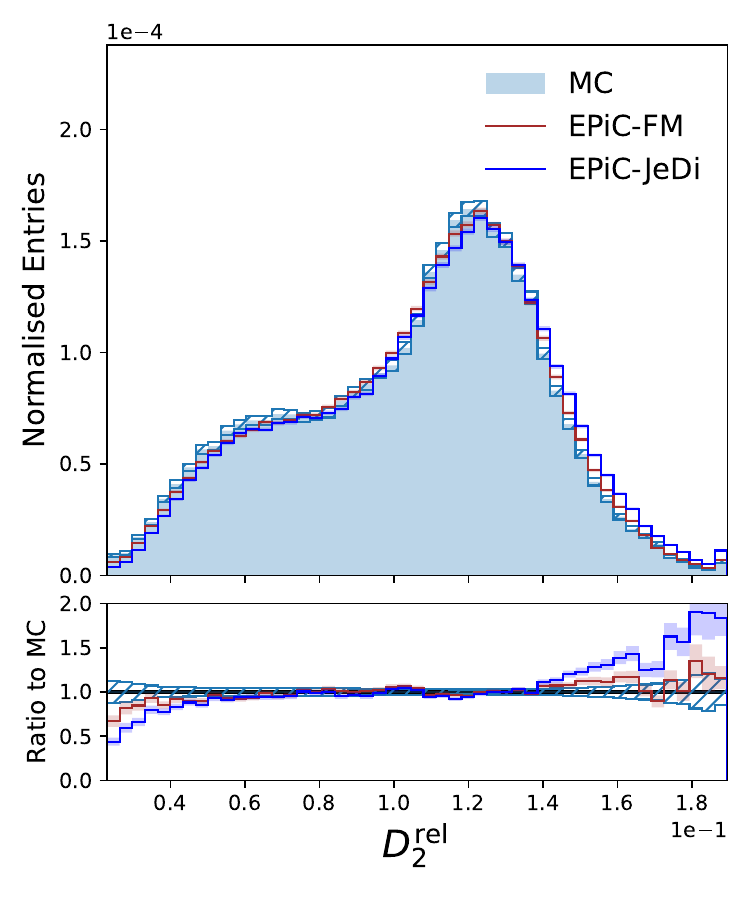}
    \end{center}
    \caption{Top jet generation with 150 constituents from Monte Carlo (light blue shade) and from conditional EPiC-FM (dark red) and conditional EPiC-JeDi (blue). First row: $p_T$ distribution of leading (left), fifth leading (middle), and twentieth leading (right) constituents. Second row: distributions of $m^{\mathrm{rel}}$ (left) and $p_T^{\mathrm{rel}}$ (right). Third row: distributions of $\tau_{21}$ (left), $\tau_{32}$ (middle), and $D_{2}$ (right). For EPiC-FM and EPiC-JeDi, a midpoint solver with 200 function evaluations is used. The error bars correspond to statistical uncertainty of the bin count.}
    \label{fig:150-top150const}
\end{figure*}

\clearpage

 \begin{table*}[t!]
    \centering
    \caption{Top jet generation with 150 constituents: Summary of performance metrics for generated jets using the Frech\'et ParticleNet Distance (FPND), the negative log-posterior (NLP) of a ParticleNet multi-model classifier (with $N\!=\!5$ classes), and the Kullback-Leibler divergence for various features with respect to JetNet-150 test dataset.}
    \label{tab:metrics150}
    \resizebox{\textwidth}{!}{%
    \begin{tabular}{llrrrrrrrrrr}
\toprule
Generation & Model& FPND & NLP & $\mathrm{KL}^{m} (\times 10^{-3})$ & $\mathrm{KL}^{p_T^{\rm const}} (\times 10^{-3})$ & $\mathrm{KL}^{\tau_{21}} (\times 10^{-3})$ & $\mathrm{KL}^{\tau_{32}} (\times 10^{-3})$ & $\mathrm{KL}^{D_2} (\times 10^{-3})$\\

\midrule
\multirow[c]{2}{*}{Conditional} & EPiC-JeDi & $0.52$ &$5.67$ & $9.10 \pm 0.79$ & $6.42 \pm 0.76$ & $14.32 \pm 1.08$ & $19.92 \pm 1.21$ & $9.40 \pm 0.88$ \\
                                & EPiC-FM & $\mathbf{0.12}$ &$\mathbf{0.12}$ & $\mathbf{4.30 \pm 0.53}$ & $\mathbf{0.84 \pm 0.02}$ & $\mathbf{9.43 \pm 0.61}$ & $\mathbf{11.22 \pm 1.02}$ & $\mathbf{4.28 \pm 0.56}$\\
\midrule
\multirow[c]{3}{*}{Unconditional} & EPiC-GAN & $0.93$ &$11.6$ & $\mathbf{6.50 \pm 0.63}$ & $2.22 \pm 0.09$ & $20.60 \pm 1.55$ & $69.64 \pm 3.30$ & $6.04 \pm 0.64$\\
                                  & EPiC-JeDi & $1.93$ &$5.70$ &$27.46 \pm 1.24$ & $6.39 \pm 0.60$ & $20.15 \pm 1.25$ & $36.50 \pm 1.81$ & $11.70 \pm 0.98$ \\
                                  & EPiC-FM & $0.18$ & $0.98$ & $12.95 \pm 0.90$ & $\mathbf{0.87 \pm 0.02}$ & $\mathbf{10.59 \pm 0.88}$ & $\mathbf{12.14 \pm 0.97}$ & $\mathbf{4.39 \pm 0.55}$\\
\bottomrule
\end{tabular}

    }
\end{table*}

We also computed the Wasserstein-1 distances for the same observables (see midpoint rows in Tab.~\ref{tab:metrics30_samplers} in \ref{app:samplers}) and found that the ones based on $m_j^\mathrm{rel}$ and $p_T^\mathrm{rel}$ are compatible with the results in Fig.~\ref{fig:30-top30} and the FPND, NLP and KL metrics in Tab.~\ref{tab:metrics30}. On the other hand, the Wasserstein metrics based on the jet substructure observables, W$_1^{\tau_{21}}$, W$_1^{\tau_{32}}$ give results that are not fully compatible with Fig.~\ref{fig:30-top30}, as explained in Sec.~\ref{sec:metrics}. There, the metrics are giving an upper hand to \epcjedi, when it is clearly performing comparably or worse than \epcfm. 

\subsection{Top jets: 150 constituents}

We now turn to the more challenging JetNet dataset with up to 150 constituents (JetNet-150). As before, we train conditional and unconditional \epcjedi and \epcfm models on the top-quark dataset. Notice that this time we have not included PC-JeDi in the conditional model comparison because the training and generation times for particle clouds with $\mathcal{O}(100)$ particles start becoming prohibitively long when using the transformer architecture. For the unconditional models we compare the performances between \epcfm, \epcjedi and \epcgan, see \ref{app:uncond_results} for the full results.
 
In Fig.~\ref{fig:150-top150const} we present the top-quark samples generated by the conditional models and compare these to the JetNet-150 training samples. These samples where produced using the same sampling schemes described in Sec.~\ref{sec:results30}. From the top row, one can notice that \epcfm (dark red) and \epcjedi (blue) are capable of producing high quality jet constituents that lead to realistic particle-level distributions approximating the truth data (blue shade). \epcfm performance matches all three distributions within uncertainties, while \epcjedi has a slight deviation from the truth level in the tails for the subleading constituents distributions. 
In general, the relative jet mass and relative transverse momentum distributions displayed in the middle row, are captured well by the two models. For instance, both models are able to reproduce the top-quark bimodal structure in the jet mass. 
Yet, there is a noticeable difference between \epcfm and \epcjedi when looking at the jet $p_T^{\rm rel}$ distribution. 
Notice that for JetNet-150, this distribution has a different shape when compared to the JetNet-30 (see Fig.~\ref{fig:30-top30}). 
When the JetNet-150 dataset was prepared, the truncation at 150 constituents only removed a small fraction (if any) of the softest constituents from each jet, leading to a (approximate) normalization constraint between jet constituents\footnote{Conversely, the much harder cut at 30 particles when preparing JetNet-30 removed a large number of hard constituents, leading to a $p_T^{\rm rel}$ distribution with a heavy tail towards lower values and a mildly truncated tail at $p_T^{\rm rel}=1$. In this case the constraint \eqref{eq:normalization} is no longer satisfied.}:
\begin{equation}\label{eq:normalization}
\sum_{i\,\in \,\rm jet} p_{T, i}^{\rm rel} \approx 1 \,. 
\end{equation}
In this case, the resulting $p_T^{\rm rel}$ distribution for jets is highly peaked at $p_T^{\rm rel}=1$ and has a tiny tail towards lower values. In the plot we can see that both generative models approximate very well the location of the mode at $p_T^{\rm rel}=1$, but it is clear that \epcfm learns more faithfully the constraint \eqref{eq:normalization}. The same result can be found for the unconditional models as shown in Fig.~\ref{fig:150-top150uncond}.  

In the last row of the figure we show the same $N$-subjetiness and energy correlation ratios described in the previous subsection. Here again, we find that both \epcfm and \epcjedi have trouble in generating the correct shapes for the bulk of the two $N$-subjetiness distributions (slightly worst than for JetNet-30), but are able to correctly capture $D_2^{\rm rel}$ within uncertainties, with a slightly better performance by \epcfm in the tails. 

Finally, in Tab~\ref{tab:metrics150} we provide the metrics for JetNet-150. We arrive at the same conclusions as for JetNet-30, that the conditional \epcfm model is outperforming \epcjedi across all metrics for the conditional models. For the unconditional models \epcgan is outperforming \epcfm only for the KL$^{m}$ metric, because of the same reasons explained in Sec.~\ref{sec:results30}.  

\begin{figure}[b]
    \centering
    \includegraphics[width=0.5\textwidth]{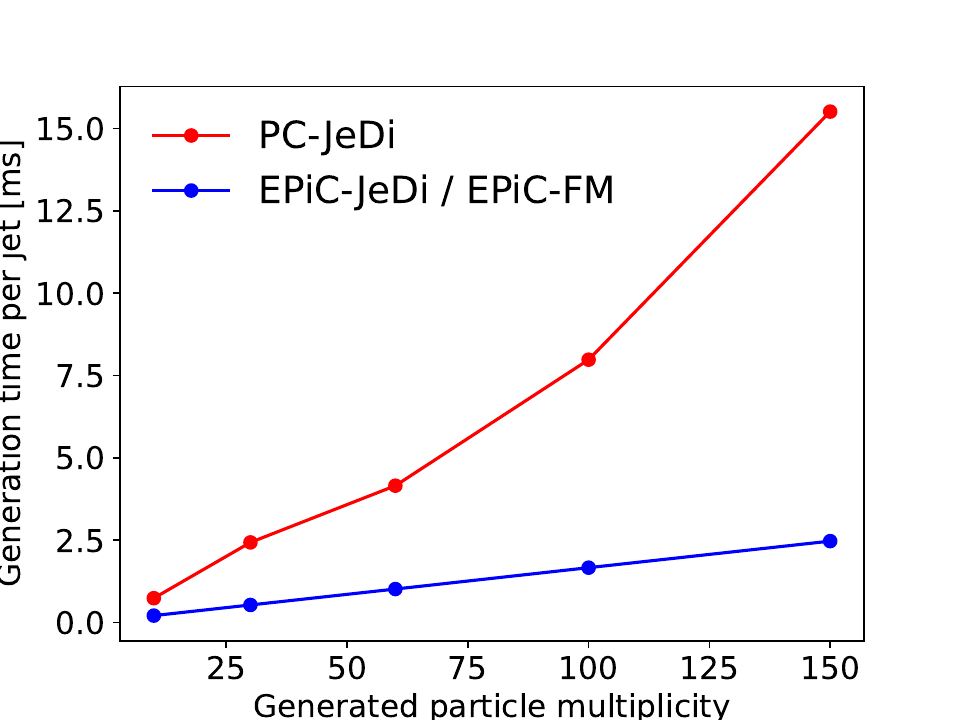}
    \vspace{0.5cm}
    \caption{Generation time per jet as a function of generated particle multiplicity. The generation time is measured on the same hardware using an \nvidia A100 40\,GB graphics card, with per-jet times calculated as the average for all test events generated. The batch size was optimised for optimal generation speed.
    As \epcjedi and \epcfm share the same architecture  the timing is representative for either approach.
    For reference, \epcgan generates jets with 10 (150) particles in a little over 1~µs (10~µs).
    }
    \label{fig:timing_study}
\end{figure}

\subsection{Timing studies}

To show the better scaling behaviour with the point clouds size of the EPiC layers, we compare the generation time of EPiC-JeDi and PC-JeDi for jets between 10 and 150 constituents.
The generation time is measured on the same hardware using an \nvidia A100 40\,GB graphics card, with per-jet times calculated as the average for 270k jets generated. 
The batch size was optimized for each generated particle multiplicity. 
This timing study was performed without retraining of the models and since EPiC-JeDi and EPiC-FM share the same model architecture, the EPiC-JeDi results are representative for either approach.

The results of this timing benchmark are shown in Fig.~\ref{fig:timing_study}.
The advantage of the EPiC layers in the EPiC network is, that they scale $\mathcal{O}(N)$ with the point cloud size $N$, while the self-attention transformer layers used in in PC-JeDi scale with $\mathcal{O}(N^2)$.
Especially at larger point size of exactly 150 samples per point, the EPiC models are $6.2\times$ faster than PC-JeDi (2.5 ms vs. 15.5 ms).
For even larger point cloud sizes, i.e. for simulating calorimeter showers with $\mathcal{O}(1000)$ points~\cite{CaloClouds, CaloClouds_2}, the advantage of \epcjedi and \epcfm would be even more dramatic.

For reference, the EPiC-GAN generates jets in about 10 $\mu$s, which is $250\times$ faster than \epcjedi and \epcfm. This is consistent as we generate samples with the midpoint ODE solver using 200 model evaluations per jet. 
The timing results of our novel EPiC models could be further improved by applying a distillation technique, such as progressive distillation~\cite{progressive_distillation} or consistency distillation~\cite{consistency_models}.
For generative models on the JetNet dataset, this has been done for the transformer diffusion models in Refs.~\cite{FPCD, PCDroid}.
Our approach to speed up the jet generation using the EPiC layers is complimentary to these distillation techniques.

\section{Conclusion}
\label{sec:conclusion}

In this work we addressed the challenge of jet generation at the LHC through deep generative models, emphasizing the representation of jets as permutation-invariant particle clouds. This led to two notable developments: i) The integration of the EPiC network into score-based diffusion models, termed \epcjedi, which aimed for a balance between computational efficiency and model accuracy when compared to its transformer-based predecessor, PC-JeDi. 
ii) We introduced the first permutation-equivariant flow-based model for particle cloud generation by leveraging the flow-matching objective in conjunction with the EPiC architecture in order to efficiently train a continuous normalizing flow, resulting in \epcfm. 

Both flow-based and diffusion-based generative modeling evolve a simple Gaussian prior into a complex one approximating the training data, utilizing continuous-time dynamics. These models aim to learn the probability density paths connecting the target data to the Gaussian distribution by training a deep neural networks to regress a suitable conditional vector field. Once trained, these models are capable of generating synthetic data by evolving noise samples towards the data distribution using deterministic or stochastic samplers. Since both generative models share many elements in common, we were able to explore different sampling methods, assess the impact of conditional versus unconditional models, and offer direct comparisons in terms of performance metrics.

In our experiments, we found that both \epcjedi and \epcfm achieve state-of-the-art performance for the JetNet-30 top-quark dataset, as well as the more challenging JetNet-150 dataset, when compared to prior leading methods like \epcgan and PC-JeDi. The flow-based \epcfm model, in particular, demonstrated a superior performance consistently across all performance metrics in both the conditional and unconditional cases. Furthermore, we also showed that both generative models gained a higher accuracy if conditional information from the jet, such as the jet mass and $p_T$ was included during training. Interestingly, the improvement in performance for \epcfm was marginal when conditioning on these variables. 

Another remarkable finding is that replacing the transformer architecture of the score-based models with the more efficient EPiC architecture not only enhances the generative fidelity, but also speeds-up the jet generation time by a factor of $6.2\times$ when generating 150 particles, reducing the need for and complementing diffusion distillation methods. 
For the envisioned generation of larger point clouds, i.e. calorimeter showers, the speed-up would be even more dramatic, as the computational efficiency of the EPiC layers scale linearly with the point cloud size, opposed to the quadratic scaling of the regular self-attention transformer layers.

With the rapid proliferation of generative models for jets in the recent literature, there is a growing need to expand the repertoire of performance metrics. This is especially important when attempting to disentangle models that might exhibit similar performance but subtle differences. To address this, we introduced two novel performance metrics for evaluating generative models for particle clouds. The first is the negative log-posterior (NLP) of a trained multi-model ParticleNet classifier evaluated on a test dataset. This new classifier metric falls into the same category as the Fr\'echet ParticleNet measure since it evaluates the (relative) performance between generative models by inspecting the full phase-space of the data. Our results showed that both metrics had good agreement. The second metric that we proposed is the Kullback-Leibler divergence KL$^\mathcal{O}$ between the generated distribution of a feature $\mathcal{O}$ and the corresponding distribution from a test dataset. We argue that this new class of metrics based on density similarities provide complementary information to the well-known Wasserstein-1 metrics W$_1^{\mathcal{O}}$ based on optimal-transport. Specifically, when evaluating the $N$-subjettiness ratios, we found that the Wasserstein metric failed to capture the true performance of some of the generative models, whereas the KL metric gave more consistent results.

\section*{Acknowledgements}
The authors would like to thank Tilman Plehn for hosting the Glühwein workshop in Heidelberg, which brought the project together with Christmas spirit. 
DAF and DSh are grateful to Sung Hak Lim for helpful discussions about the theory of diffusion and flow-matching models. 
EB, CE, and GK thank the other members of the joint DESY-UHH generative group for many useful discussions of generative ML models.
TG, ML, GQ, JR, and DSe would like to acknowledge funding through the SNSF Sinergia grant CRSII$5\_193716$ called ``Robust Deep Density Models for High-Energy Particle Physics and Solar Flare Analysis (RODEM)'', and the SNSF project grant 200020\_212127 called ``At the two upgrade frontiers: machine learning and the ITk Pixel detector''.
ML would also like to acknowledge the funding acquired through the Swiss Government Excellence Scholarships for Foreign Scholars.
EB is funded by a scholarship of the Friedrich Naumann Foundation for Freedom and by the German Federal Ministry of Science and Research (BMBF) via Verbundprojekts 05H2018 - R\&D COMPUTING (Pilotmaßnahme ErUM-Data) Innovative Digitale Technologien für die Erforschung von Universum und Materie.
EB, CE, and GK acknowledge support by the Deutsche Forschungsgemeinschaft under Germany’s Excellence Strategy – EXC 2121  Quantum Universe – 390833306 
and via the KISS consortium (05D23GU4) funded by the German Federal Ministry of Education and Research BMBF in the ErUM-Data action plan.
DAF and DSh are supported by DOE grant DOE-SC0010008. 
This research was supported in part through the Maxwell computational resources operated at Deutsches Elektronen-Synchrotron DESY, Hamburg, Germany.

\printbibliography

\appendix
\section{Hyperparameters}
\label{app:hyperparams}

For most hyperparameters in \epcjedi and \epcfm, we follow the choices in \pcjedi~\cite{PCJedi} and \epcgan~\cite{EPiCGAN} respectively. An overview of the hyperparameters can be found in Table~\ref{tab:hyperparameters}.
In all cases, the weights of the model after the final training epoch are used for generation, and no epoch picking based on evaluation metrics is performed.

\begin{table}[H]
\centering
\caption{Hyperparameter choices for \epcjedi and \epcfm.}
\label{tab:hyperparameters}
\begin{tabular}{lr}
\hline
 Hyperparameter & Value \\
 \hline
 \hline
    EPiC layers  & 6 \\
    EPiC global dimensionality  & 10 \\
    Hidden dimensionality & 128 \\
    Activation function & LeakyReLU($0.01$) \\
    Adam-W~\cite{AdamW} learning rate & $10^{-3}$ \\
    Learning rate scheduling & Cosine with warm-up \\
    Warm-up epochs & 1,000 \\
    Batch size &  1,024   \\
    Training epochs & 10,000 \\
\hline
    Model weights & $\sim 560,000$ \\
\hline
    Training events & $\sim 110,000$ \\
    Test events & $\sim 27,000$ \\
\hline
\end{tabular}

\end{table}

\section{Multi-model classifier metric}
\label{app:classifier}

In this appendix we provide details on the multi-model classifier training an architecture. From each trained generative model we sampled around $220,000$ jets with variable sizes. The data was then split into training, validation and testing datasets following $0.6/0.1/0.3$ fractions.  

We used the relative constituent positions $x=(\Delta\eta,\Delta\phi)$ for the node coordinates and $(p_T^{\rm rel}, E^{\rm rel} , \Delta R)$  for the node features. Here $E^{\rm rel}=E^{\rm const}/E^{\rm jet}$ is the relative constituent energy and $\Delta R=\sqrt{\Delta\eta^2+\Delta\phi^2}$ is the distance between the constituent and the jet axis. The node features for the training datasets where further preprocessed by normalising by their mean and standard deviation. 

For the classifier architecture we used a ParticleNet-Lite network \cite{ParticleNet}  with two Edge-Conv blocks, each with $k=8$ nearest neighbours and with $C\!=\!(32,32,32)$ and $C\!=\!(64,64,64)$ channels, respectively. After the EdgeConv, the output was passed through a channel-wise global average pooling operation and fed into a fully connected layer with $256$ hidden units with ReLU activation function, batch-normalisation layer and a dropout probability of $0.1$. In the final stage, we applied the softmax function on the final $N$-class output. For the training we used a batch-size of $2,048$, the Adam optimizer \cite{Adam} with a learning rate of $10^{-3}$ and cosine annealing. We trained the classifier for a total of 5,000 epochs and selected the best classifier based on the lowest validation loss.

\section{Additional Results for Unconditional Generation}
\label{app:uncond_results}

\begin{figure*}[h]
    \begin{center}
    \includegraphics[width=.32\textwidth]{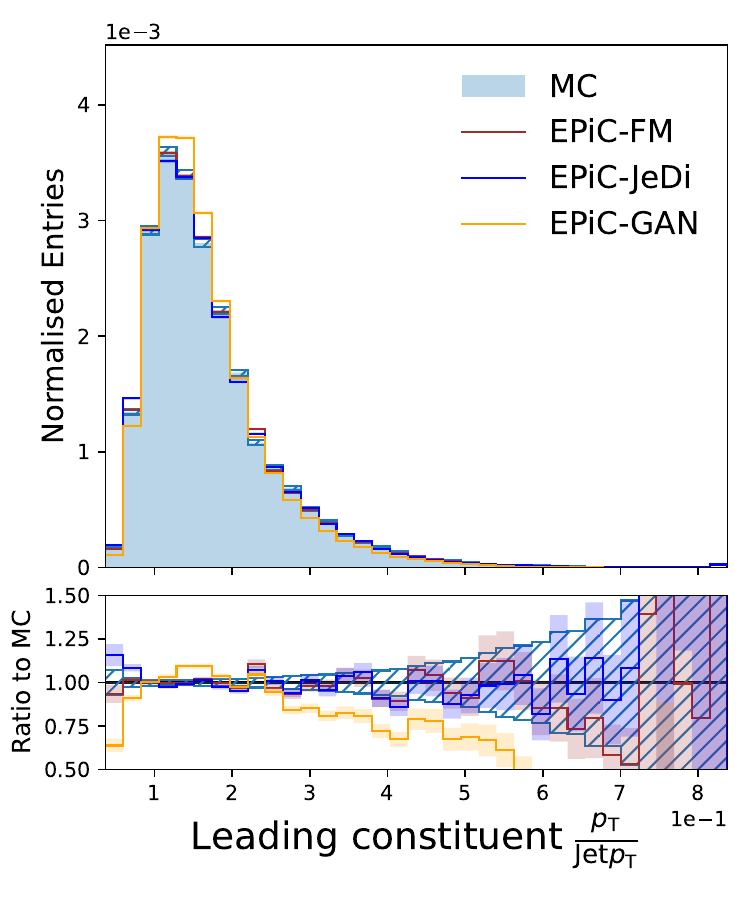}
    \includegraphics[width=.32\textwidth]{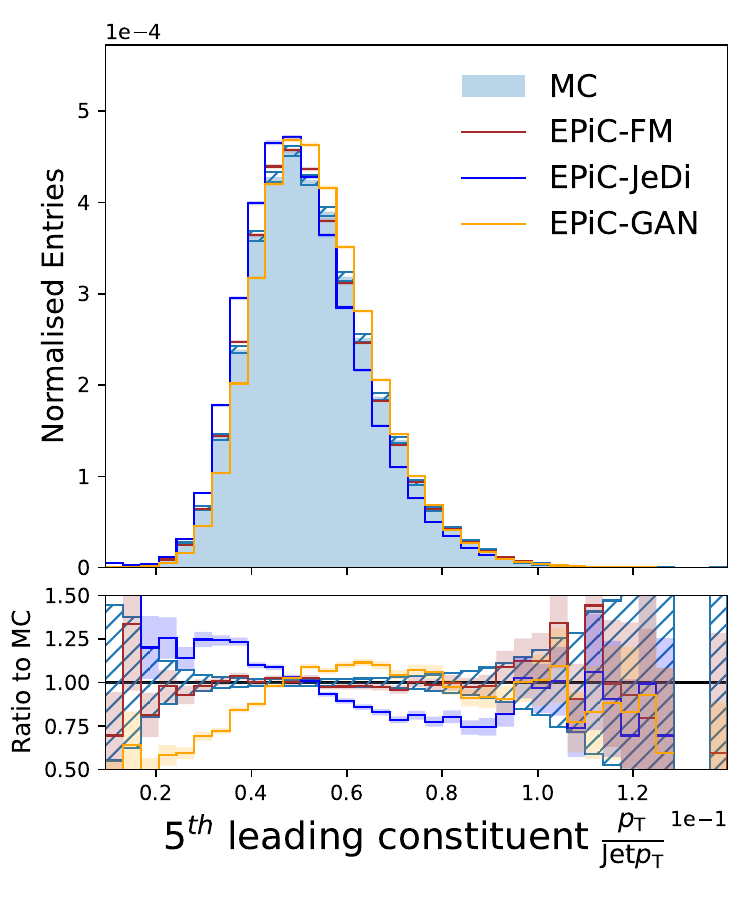}
    \includegraphics[width=.32\textwidth]{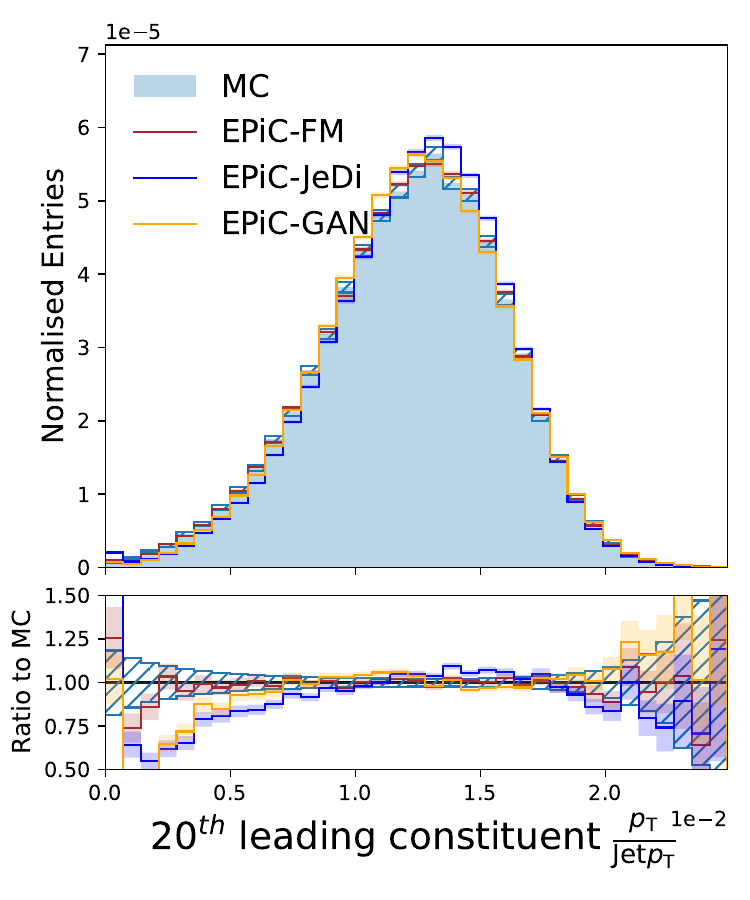}
    \end{center}
    \begin{center}
    \includegraphics[width=.32\textwidth]{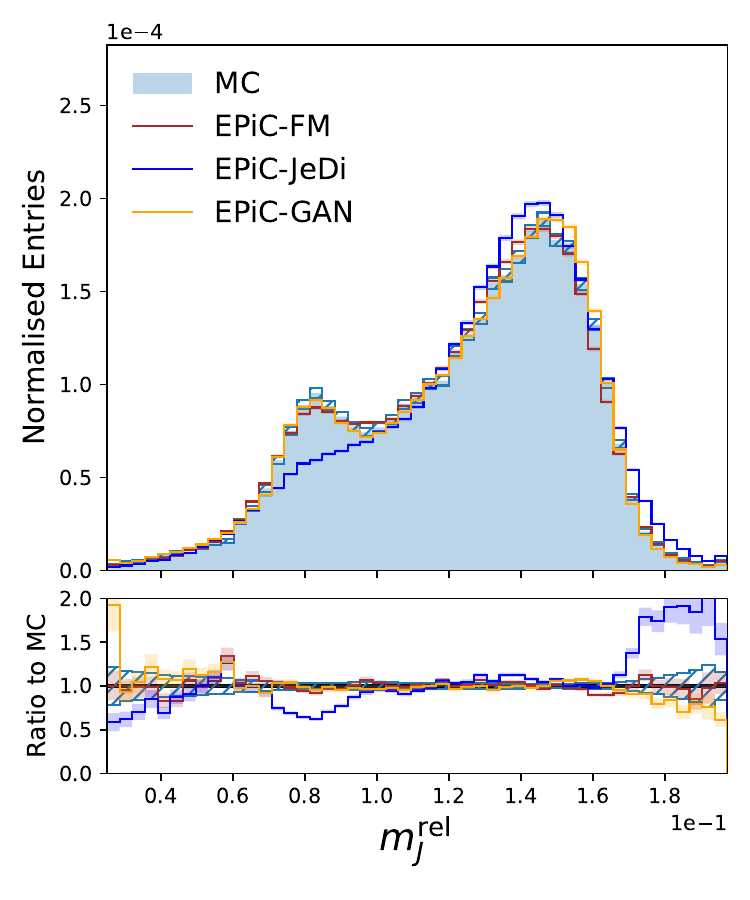}
    \includegraphics[width=.32\textwidth]{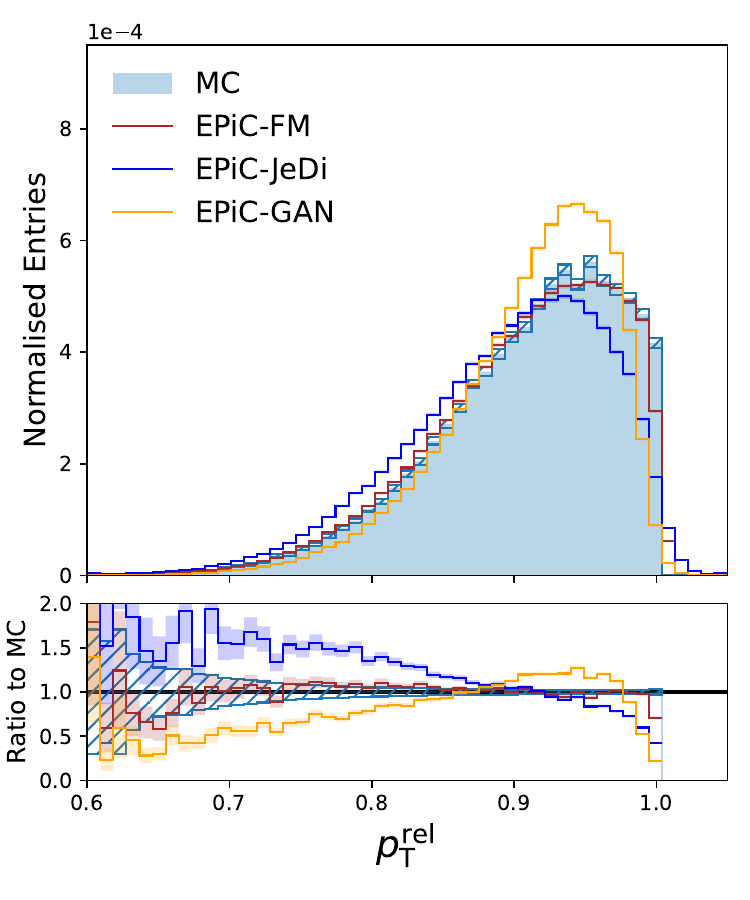}
    \end{center}
    \begin{center}
    \includegraphics[width=.32\textwidth]{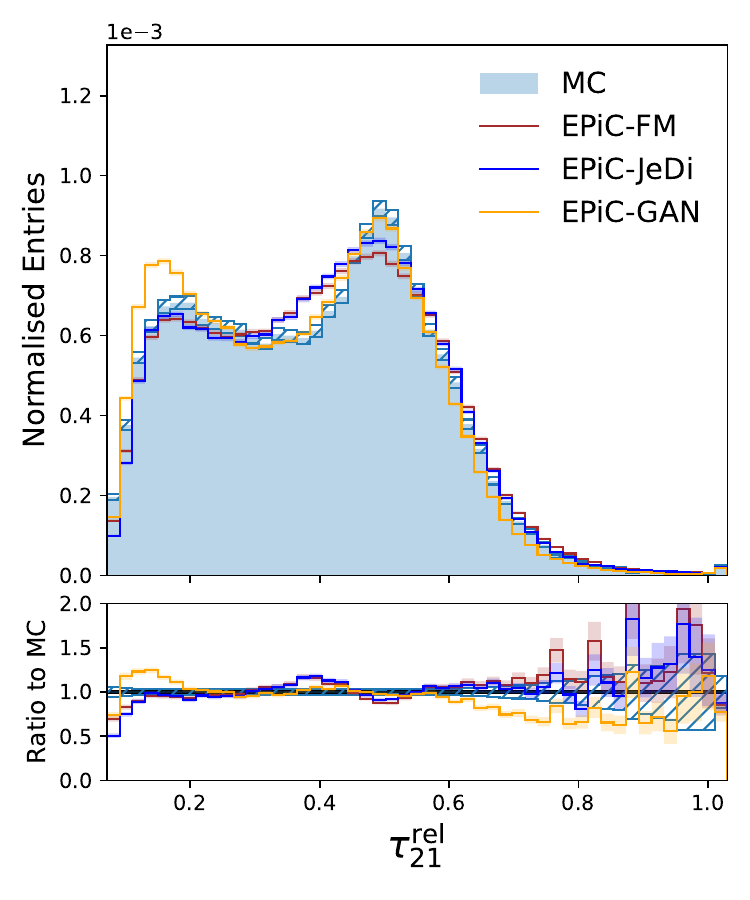}
    \includegraphics[width=.32\textwidth]{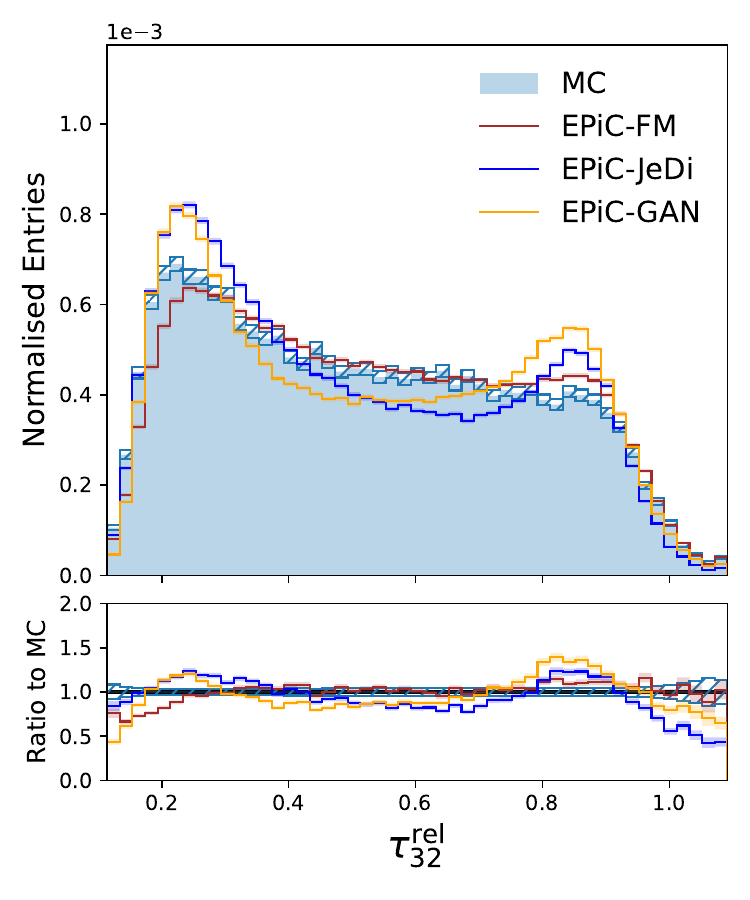}
    \includegraphics[width=.32\textwidth]{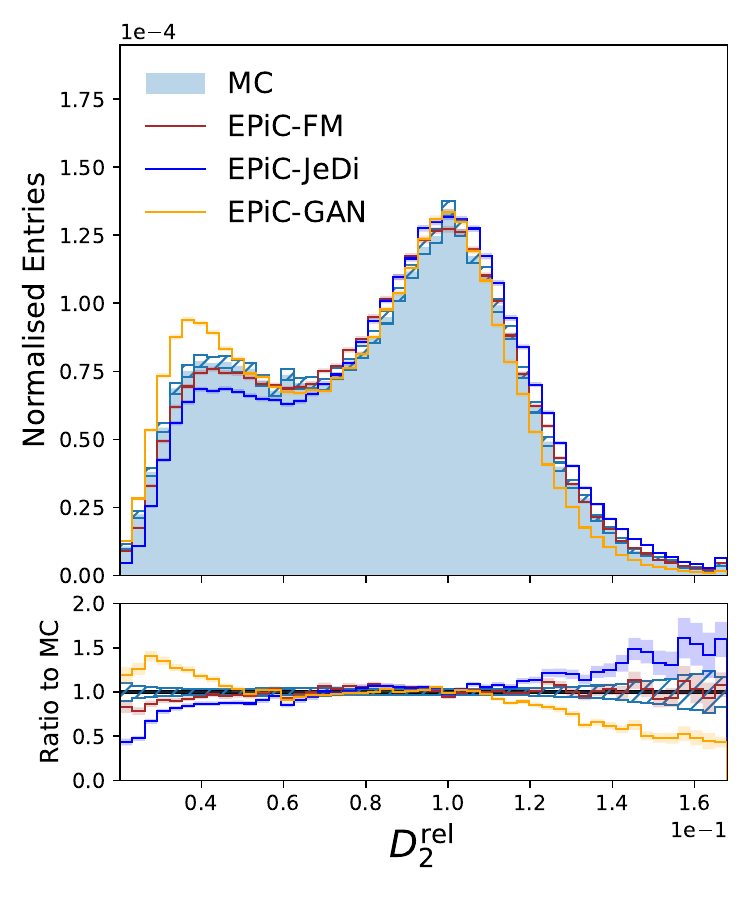}
    \end{center}
    \caption{Top jet generation with 30 constituents and unconditional models: $p_T$ distribution of leading (left), fifth leading (middle), and twentieth leading (right) constituents of top jets generated with Monte Carlo (light blue shade), EPiC-FM (dark red), EPiC-JeDi (blue) and \epcgan (yellow). For EPiC-FM and EPiC-JeDi, a midpoint solver with 200 function evaluations is used. The error bars correspond to statistical uncertainty of the bin count.}
    \label{fig:30-top30uncond}
\end{figure*}

\begin{figure*}[h]
    \begin{center}
    \includegraphics[width=.32\textwidth]{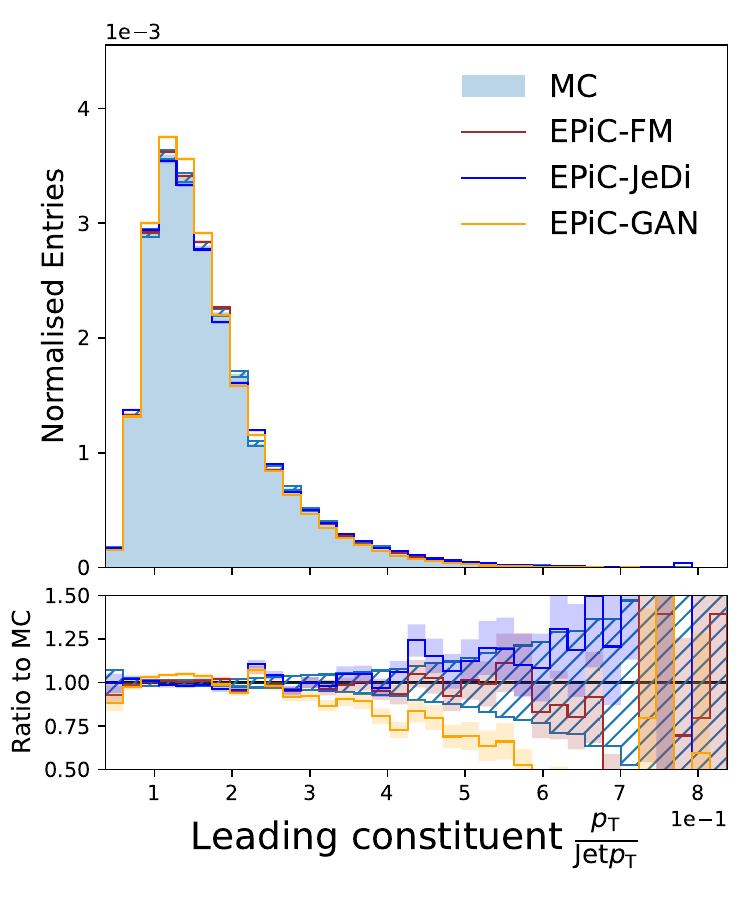}
    \includegraphics[width=.32\textwidth]{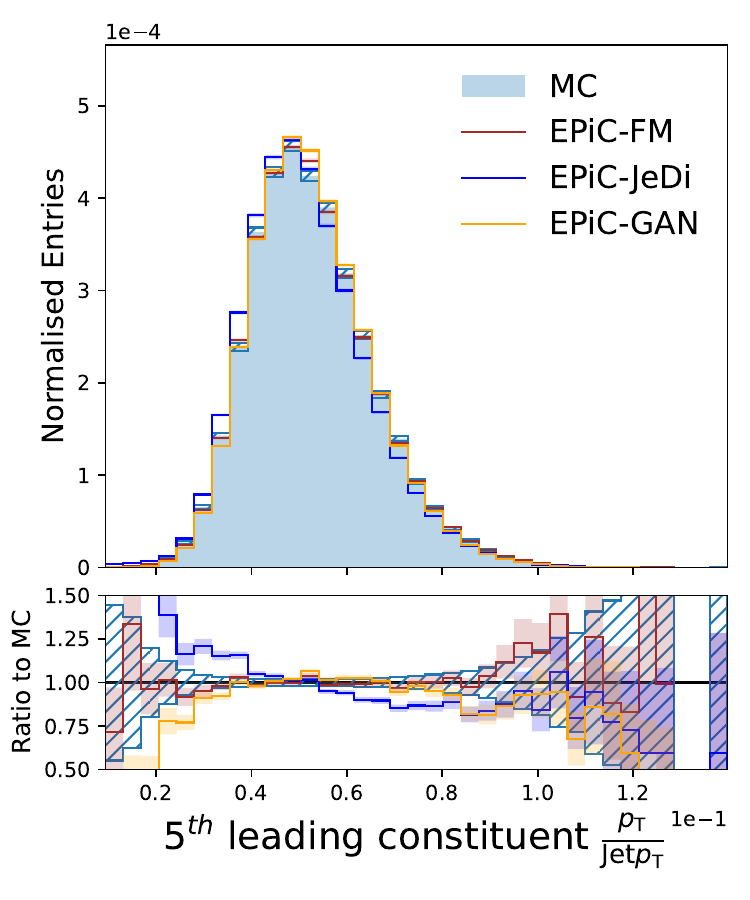}
    \includegraphics[width=.32\textwidth]{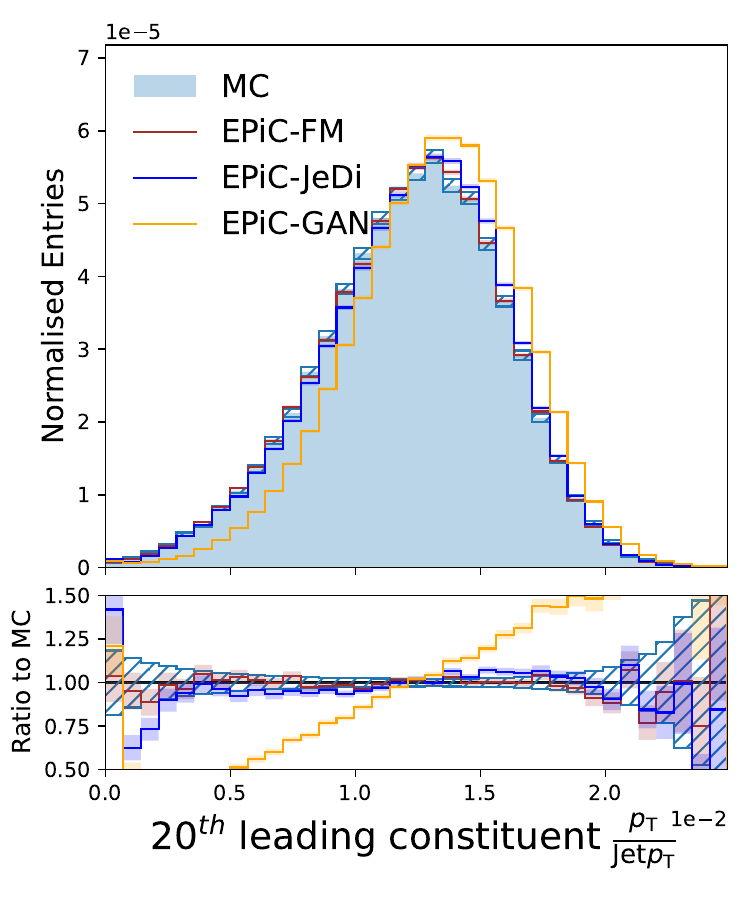}
    \end{center}
    \begin{center}
    \includegraphics[width=.32\textwidth]{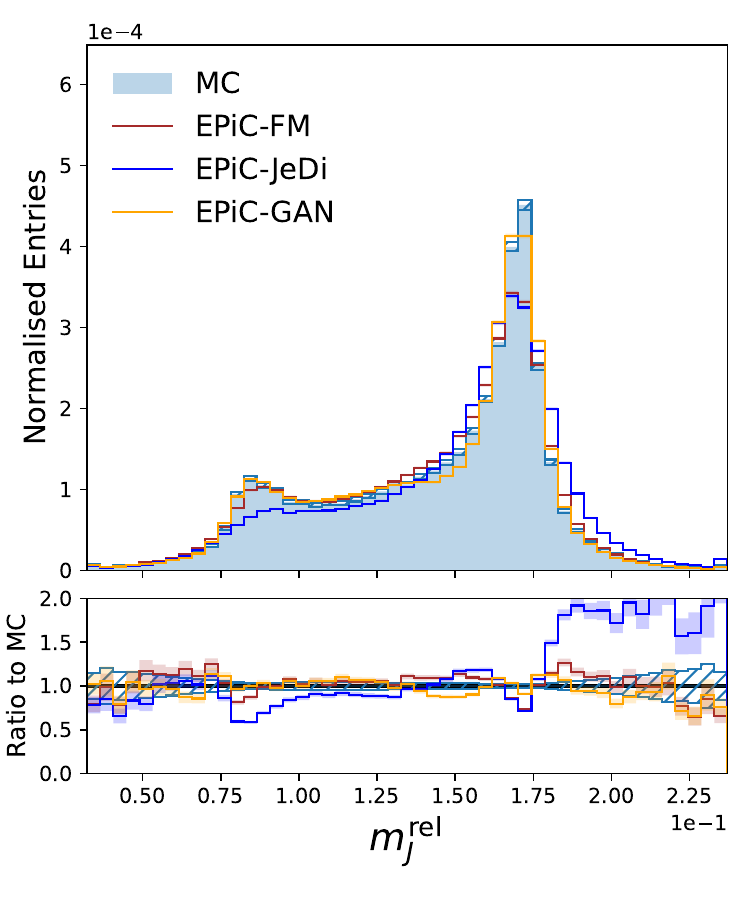}
    \includegraphics[width=.32\textwidth]{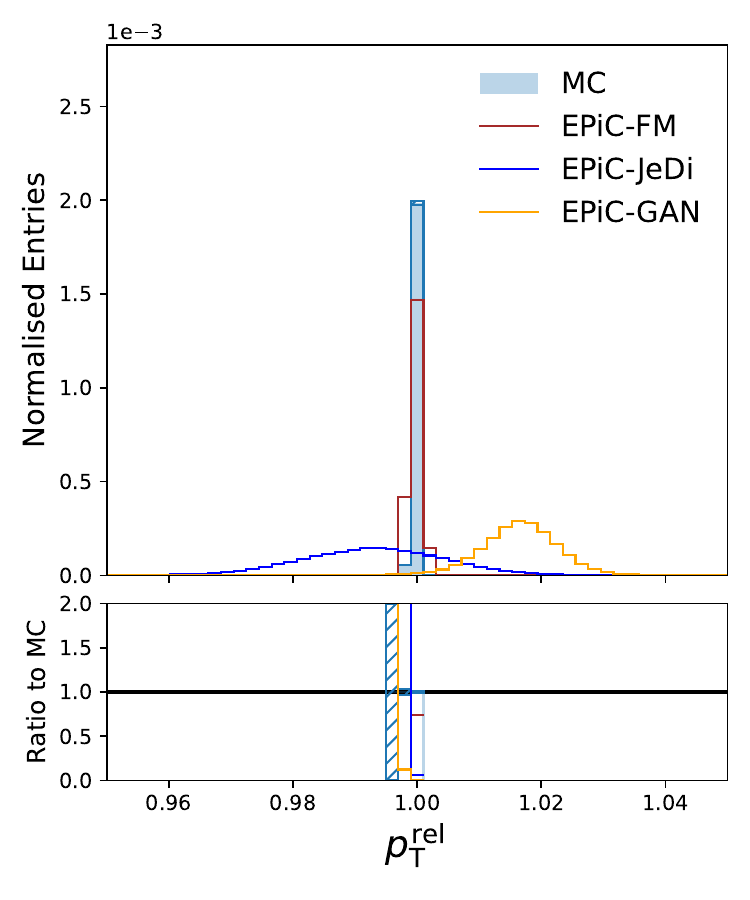}
    \end{center}
    \begin{center}
    \includegraphics[width=.32\textwidth]{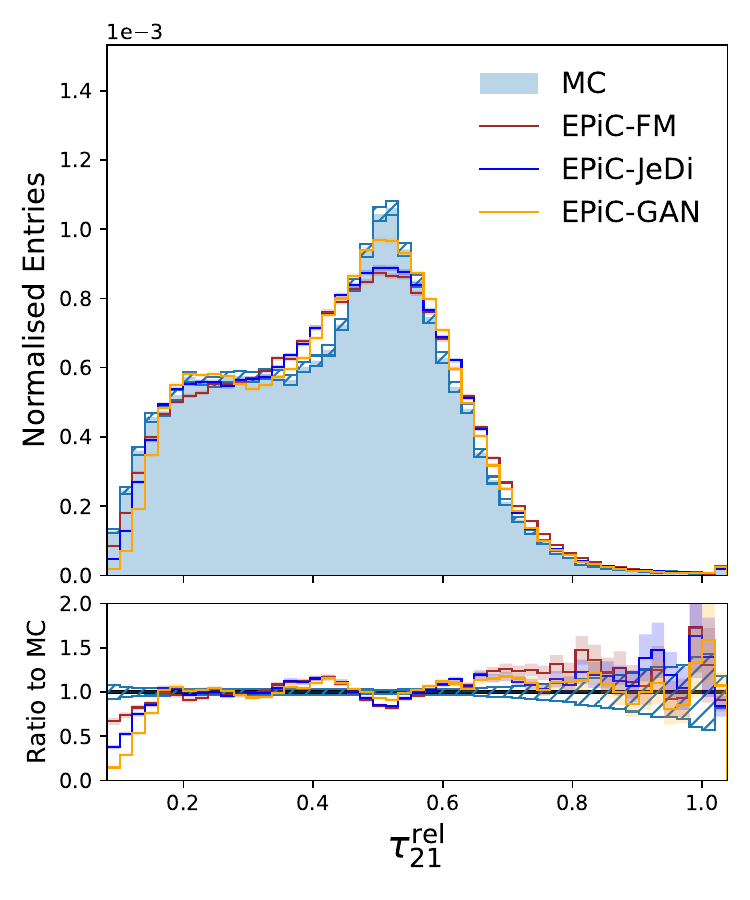}
    \includegraphics[width=.32\textwidth]{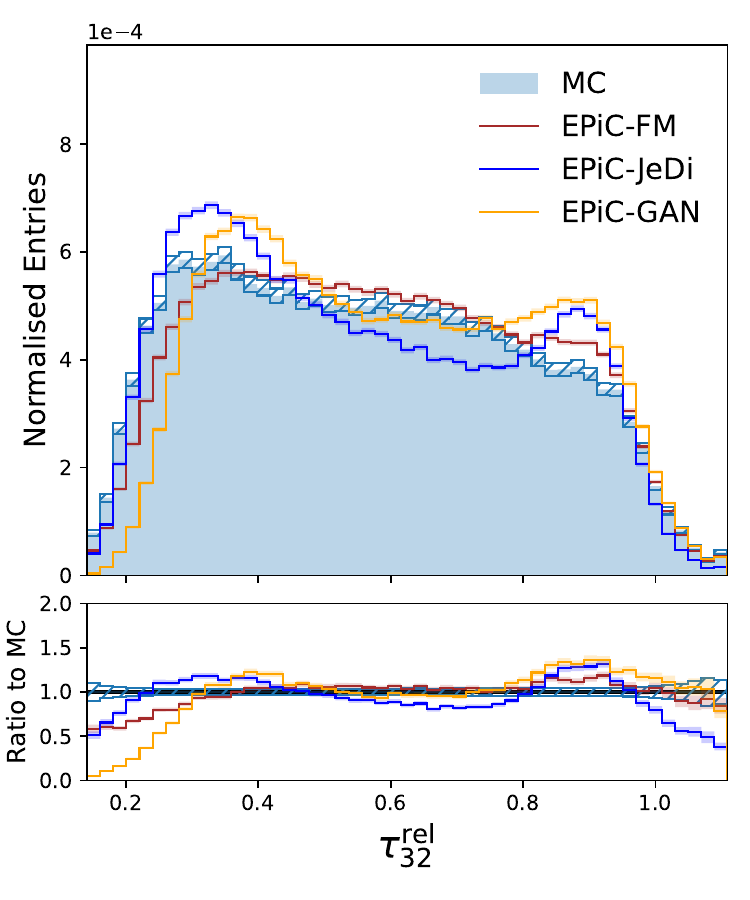}
    \includegraphics[width=.32\textwidth]{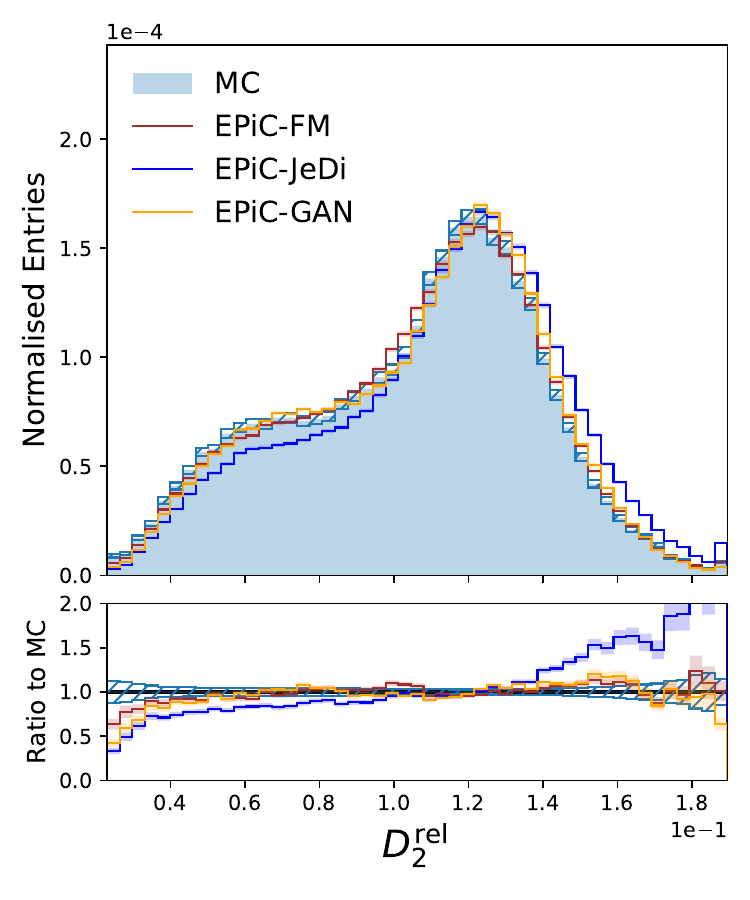}
    \end{center}
    \caption{Top jet generation with 150 constituents and unconditional models: $p_T$ distribution of leading (left), fifth leading (middle), and twentieth leading (right) constituents of top jets generated with Monte Carlo (light blue shade), EPiC-FM (dark red), EPiC-JeDi (blue) and \epcgan (yellow). For EPiC-FM and EPiC-JeDi, a midpoint solver with 200 function evaluations is used. The error bars correspond to statistical uncertainty of the bin count.}
    \label{fig:150-top150uncond}
\end{figure*}

In Fig.~\ref{fig:30-top30uncond}
we present results for the three unconditional models \epcjedi (blue), \epcfm (dark red) and \epcgan (yellow) trained on the top-quark JetNet-30 datasets (shaded blue). 
In Fig.~\ref{fig:150-top150uncond} we present results for the unconditional models \epcjedi (blue), \epcfm (dark red) and \epcgan (yellow) trained on the top-quark JetNet-150 datasets (shaded blue). 
In both figures we compare the unconditional \epcjedi and \epcfm models to the unconditional pre-trained EPiC-GAN from Ref.~\cite{EPiCGAN}.
A similar picture emerges as seen for the conditional models in Sec.~\ref{sec:results}: The \epcfm models outperform both \epcjedi and \epcjedi in all presented observables.

\begin{table*}[ht]
    \centering
    \caption{Top jet generation with 30 constituents and various samplers: Summary of performance for generated jets using the metrics introduced in Ref.~\cite{MPGAN}.}
    \label{tab:metrics30_samplers}
    \resizebox{\textwidth}{!}{%
    \begin{tabular}{lllrrrrrrr}
\toprule
Generation & Model & Sampler &  \textbf{FPND} & $\mathbf{\mathrm{W}_1^{m} (\times 10{^{-4}})}$ & $\mathbf{\mathrm{W}_1^{p_T} (\times 10{^{-4}})}$ & $\mathbf{\mathrm{W}_1^{EFP} (\times 10{^{-5}})}$ & $\mathbf{\mathrm{W}_1^{\tau_{21}} (\times 10{^{-3}})}$ & $\mathbf{\mathrm{W}_1^{\tau_{32}} (\times 10{^{-3}})}$ & $\mathbf{\mathrm{W}_1^{D_2} (\times 10{^{-3}})}$ \\
\midrule
\multirow[c]{5}{*}{Conditional}
& \multirow[c]{3}{*}{EPiC-JeDi} & EM~(SDE) & $0.29$ & $16.96 \pm 2.00$ & $5.32 \pm 1.10$ & $3.47 \pm 0.38$ & $7.84 \pm 0.77$ & $26.36 \pm 1.41$ & $\mathbf{0.81 \pm 0.07}$ \\
&                               & Midpoint & $0.42$ & $8.29 \pm 1.20$ & $14.67 \pm 1.38$ & $1.76 \pm 0.22$ & $\mathbf{5.09 \pm 0.43}$ & $14.19 \pm 0.83$ & $1.35 \pm 0.22$ \\
&                               & Euler & $0.39$ & $8.65 \pm 1.14$ & $14.65 \pm 1.68$ & $1.79 \pm 0.25$ & $5.60 \pm 0.46$ & $\mathbf{13.83 \pm 1.11}$ & $1.37 \pm 0.17$ \\                      
\cline{2-10}
& \multirow[c]{2}{*}{EPiC-FM} & Midpoint & $\mathbf{0.11}$ & $\mathbf{5.12 \pm 1.18}$ & $\mathbf{3.36 \pm 0.98}$ & $\mathbf{1.10 \pm 0.26}$ & $7.54 \pm 0.84$ & $16.33 \pm 1.21$ & $0.97 \pm 0.17$ \\
&                             & Euler & $0.19$ & $13.26 \pm 1.85$ & $10.95 \pm 1.40$ & $3.11 \pm 0.35$ & $10.54 \pm 1.12$ & $18.72 \pm 1.36$ & $1.13 \pm 0.11$ \\
\midrule

\multirow[c]{5}{*}{Unconditional}
& \multirow[c]{3}{*}{EPiC-JeDi} & EM~(SDE) & $0.77$ & $16.92 \pm 1.36$ & $14.52 \pm 1.73$ & $2.88 \pm 0.20$ & $12.62 \pm 0.82$ & $\mathbf{12.09 \pm 0.75}$ & $2.19 \pm 0.18$ \\
&                               & Midpoint & $1.63$ & $37.54 \pm 1.91$ & $33.57 \pm 1.48$ & $8.08 \pm 0.40$ & $\mathbf{7.71 \pm 0.99}$ & $15.73 \pm 1.17$ & $3.69 \pm 0.19$ \\
&                               & Euler & $1.64$ & $37.10 \pm 1.72$ & $32.63 \pm 1.59$ & $8.33 \pm 0.44$ & $8.56 \pm 0.87$ & $14.29 \pm 0.86$ & $3.86 \pm 0.18$ \\
\cline{2-10}
& \multirow[c]{2}{*}{EPiC-FM} & Midpoint & $\mathbf{0.14}$ & $\mathbf{7.69 \pm 0.97}$ & $\mathbf{3.39 \pm 0.98}$ & $\mathbf{1.45 \pm 0.30}$ & $7.77 \pm 0.80$ & $14.97 \pm 1.39$ & $\mathbf{0.94 \pm 0.17}$ \\
&                             & Euler & $0.39$ & $30.16 \pm 1.78$ & $17.55 \pm 1.49$ & $6.43 \pm 0.42$ & $8.41 \pm 0.72$ & $23.53 \pm 1.37$ & $1.40 \pm 0.10$ \\
\bottomrule
\end{tabular}
    }
\end{table*}

\begin{table*}[ht]
    \centering
    \caption{Top jet generation with 150 constituents and various samplers: Summary of performance for generated jets using the metrics introduced in Ref.~\cite{MPGAN}.}
    \label{tab:metrics150_samplers}
    \resizebox{\textwidth}{!}{%
    \begin{tabular}{lllrrrrrrr}
\toprule
Generation & Model & Sampler &  \textbf{FPND} & $\mathbf{\mathrm{W}_1^{m} (\times 10{^{-4}})}$ & $\mathbf{\mathrm{W}_1^{p_T} (\times 10{^{-4}})}$ & $\mathbf{\mathrm{W}_1^{EFP} (\times 10{^{-5}})}$ & $\mathbf{\mathrm{W}_1^{\tau_{21}} (\times 10{^{-3}})}$ & $\mathbf{\mathrm{W}_1^{\tau_{32}} (\times 10{^{-3}})}$ & $\mathbf{\mathrm{W}_1^{D_2} (\times 10{^{-3}})}$ \\
\midrule
\multirow[c]{5}{*}{Conditional}
& \multirow[c]{3}{*}{EPiC-JeDi} & EM (SDE) & $0.26$ & $10.12 \pm 2.05$ & $6.46 \pm 0.78$ & $5.77 \pm 0.81$ & $7.60 \pm 0.42$ & $31.34 \pm 1.52$ & $1.97 \pm 0.23$ \\
&                               & Midpoint & $0.52$ & $6.61 \pm 1.05$ & $18.89 \pm 1.25$ & $4.78 \pm 0.62$ & $\mathbf{7.51 \pm 0.43}$ & $21.15 \pm 1.25$ & $3.13 \pm 0.23$ \\
&                               & Euler & $0.47$ & $6.77 \pm 1.55$ & $18.80 \pm 1.29$ & $4.97 \pm 0.71$ & $8.73 \pm 0.58$ & $21.77 \pm 1.29$ & $3.39 \pm 0.16$ \\                     
\cline{2-10}
& \multirow[c]{2}{*}{EPiC-FM} & Midpoint & $\mathbf{0.12}$ & $\mathbf{3.74 \pm 0.89}$ & $\mathbf{3.14 \pm 1.07}$ & $\mathbf{2.30 \pm 0.42}$ & $8.51 \pm 0.98$ & $\mathbf{20.67 \pm 1.33}$ & $1.47 \pm 0.19$ \\
&                             & Euler & $0.15$ & $4.08 \pm 0.88$ & $14.24 \pm 1.18$ & $2.38 \pm 0.49$ & $8.92 \pm 0.87$ & $22.54 \pm 1.04$ & $\mathbf{0.65 \pm 0.12}$ \\
\midrule

\multirow[c]{5}{*}{Unconditional}
& \multirow[c]{3}{*}{EPiC-JeDi} & EM (SDE) & $0.52$ & $31.37 \pm 2.53$ & $8.46 \pm 1.29$ & $13.79 \pm 0.91$ & $8.82 \pm 0.62$ & $21.56 \pm 1.65$ & $3.30 \pm 0.19$ \\
&                               & Midpoint & $1.93$ & $66.07 \pm 2.05$ & $35.04 \pm 1.51$ & $27.84 \pm 0.86$ & $\mathbf{8.75 \pm 0.97}$ & $11.67 \pm 0.60$ & $6.24 \pm 0.26$ \\
&                               & Euler & $1.90$ & $66.85 \pm 2.14$ & $35.67 \pm 1.53$ & $28.03 \pm 0.97$ & $9.90 \pm 0.89$ & $\mathbf{11.40 \pm 0.82}$ & $6.30 \pm 0.19$ \\
\cline{2-10}
& \multirow[c]{2}{*}{EPiC-FM} & Midpoint & $\mathbf{0.18}$ & $\mathbf{10.77 \pm 1.12}$ & $\mathbf{3.25 \pm 0.89}$ & $\mathbf{4.03 \pm 0.37}$ & $9.37 \pm 0.74$ & $19.85 \pm 1.29$ & $\mathbf{1.11 \pm 0.18}$ \\
&                             & Euler & $0.47$ & $31.86 \pm 2.05$ & $21.79 \pm 1.45$ & $10.66 \pm 0.78$ & $9.65 \pm 0.91$ & $28.16 \pm 1.43$ & $1.52 \pm 0.15$ \\
\bottomrule
\end{tabular}
    }
\end{table*}

\section{Sampler Comparison}
\label{app:samplers}

In Tab.~\ref{tab:metrics30_samplers} and Tab.~\ref{tab:metrics150_samplers} we present results for various metrics for the \epcjedi and \epcfm models and various ODE and SDE solvers on the top-30 and top-150 dataset, respectively.
The Wasserstein-1 scores $W_1^\mathcal{O}$ for observables $\mathcal{O}$ were introduced in Ref.~\cite{MPGAN} and used in various subsequent publications~\cite{JetFlow, EPiCGAN, GAPT, PCJedi, FPCD, PCDroid}.
Note, that we calculated these scores with higher statistics than previously performed so increase the accuracy of the scores. 

Here, we compare the Euler-Maruyama~(EM), midpoint, and Euler solvers for \epcjedi as well as the midpoint and Euler solvers for \epcfm. For all solvers, samples were generated with 200 function evaluations. We find that for virtually all metrics, we find best performance with the midpoint ODE solver, which is why we have used only this solver for presenting the performance of our EPiC models in Sec.~\ref{sec:results}.

\end{document}